\documentclass[twocolumns]{aa}
\usepackage{graphicx}
\usepackage{comment}
\usepackage{url}
\usepackage[breaklinks,colorlinks,citecolor=blue]{hyperref}
\usepackage{lscape}
\usepackage{txfonts}
\begin{document}

   \title{A 3D view of multiple populations kinematics in Galactic globular clusters}

   %\subtitle{I. Overviewing the $\kappa$-mechanism}

   \author{E. Dalessandro\inst{1}
   \and
   M. Cadelano\inst{1,2}
   \and
   A. Della Croce\inst{1,2}
   \and
   F. I. Aros\inst{3}
   \and
   E. B. White\inst{3}
   \and 
   E. Vesperini\inst{3}
   \and
   C. Fanelli\inst{1}
   \and
   F. R. Ferraro\inst{1,2}
   \and
   B. Lanzoni\inst{1,2}
   \and
   S. Leanza\inst{1,2}
   \and
   L. Origlia\inst{1}
}

\institute{INAF -- Astrophysics and Space Science Observatory Bologna, Via Gobetti 93/3 I-40129 Bologna, Italy\\
\email{emanuele.dalessandro@inaf.it}
\and
Dipartimento di Fisica e Astronomia, Via Gobetti 93/2 I-40129 Bologna, Italy
\and
Department of Astronomy, Indiana University, Swain West, 727 E. 3rd Street, IN 47405 Bloomington, USA
 }

   \date{Received June 10, 2024; accepted September 02, 2024}

% \abstract{}{}{}{}{} 
% 5 {} token are mandatory
 
  \abstract{
We present the first 3D kinematic analysis of multiple stellar populations (MPs) in a representative sample of 16 Galactic globular clusters (GCs).
For each GC in the sample we studied the MP line-of-sight, plane-of-the-sky and 3D rotation as well as 
the velocity distribution anisotropy. 
The differences between first- (FP) and second-population (SP) kinematic patterns were constrained by means of parameters specifically defined
to provide a global measure of the relevant physical quantities and to enable a meaningful comparison among different clusters. 
Our analysis provides the first observational description of the MP
kinematic properties and of the path they follow during the long-term dynamical evolution. 
In particular, we find evidence of differences between the rotation of MPs along 
all velocity components with the SP preferentially rotating faster than the FP.
The difference between the rotation strength of MPs is anti-correlated with the cluster dynamical age.
We observe also that FPs are characterized by isotropic velocity distributions at any dynamical age probed by our sample. On the contrary, the velocity distribution of SP stars is found to be radially anisotropic in dynamically young clusters and isotropic at later evolutionary stages.
The comparison with a set of numerical simulations shows that these observational results are consistent 
with the long-term evolution of clusters forming with an initially more centrally concentrated and more rapidly rotating SP subsystem.
We discuss the possible implications these findings have on our understanding of MP formation 
and early evolution. 
}

   \keywords{globular clusters: general
-- Stars: Hertzsprung-Russell and C-M diagrams - kinematics and dynamics - abundances -- techniques: photometry - spectroscopy - astrometry}

\titlerunning{3D kinematics of MPs}
\authorrunning{Dalessandro et al.}
   \maketitle

%%%%%%%%%%%%%%%%%%%%%%%%%%%%%%%%%%%%%%%%%%%%%%%%%%

%%%%%%%%%%%%%%%%% BODY OF PAPER %%%%%%%%%%%%%%%%%%

\section{Introduction}
Globular clusters (GCs) exhibit intrinsic star-to-star variations in their light 
element content. In fact, while some GC stars have the same light-element abundances 
as field stars with the same metallicity (first population/generation--FP), 
others show enhanced N and Na along with depleted C and O abundances 
(second population/generation--SP). The manifestation of such light-element 
inhomogeneities is referred to as multiple populations (MPs -- see \citealt{bastianlardo18,gratton19} for a review of the subject).

Light-element abundance variations can have an impact on both the stellar structures (as in the case of He for example) and atmospheres (as for Na, O, C and N) and they can therefore produce a broadening or splitting of different evolutionary sequences
in color-magnitude-diagrams (CMDs) when appropriate filter combinations are
used \citep{piotto07,sbordone11,dalessandro11,monelli13,piotto15,nied17,cadelano23}.

It is well established now that the MP phenomenon is (almost) ubiquitous among massive stellar clusters. In fact, it has been shown that nearly all massive ($>10^4 M_{\odot}$; 
e.g., \citealt{dalessandro14,piotto15,milone17,bragaglia17}) 
and relatively old (>1.5-2 Gyr; \citealt{martocchia18a,cadelano22}) GCs host MPs. 
In addition, MPs are observed in GC systems in any environment where they have been properly searched for. They are regularly found
in the Magellanic Clouds stellar clusters \citep{mucciarelli09,dalessandro16}, 
in GCs in dwarf galaxies such as Fornax \citep{larsen12,larsen18} and Sagittarius (e.g., \citealt{sills19}), in the M31 GC system \citep{schiavon13,nardiello18} and there
are strong indications (though indirect) that they are ubiquitous in stellar clusters in massive elliptical galaxies (e.g., \citealt{chung11}).

MPs are believed to form during the very early epochs of GC formation and evolution ($\sim10-100$ Myr; 
see \citealt{martocchia18b,nardiello15,saracino20} for direct observational constraints) and a number of scenarios have been proposed over the years to describe the sequence of physical events and mechanisms involved in their formation. We can schematically group them in two main categories. One includes multi-epoch formation models, which predict that MPs form during multiple (at least two) events of star formation and they typically invoke self-enrichment processes, in which the SP forms out of the ejecta of relatively massive FP stars \citep[e.g.][]{decressin07,dercole08,demink09,dantona16}. The second category includes models where MPs form simultaneously and SPs are able to accrete gas eventually during their pre-main sequence phases (e.g., \citealt{bastian13,gieles18}).
However, independently on the specific differences, all models proposed so far have their own caveats and face serious problems to reproduce some or all the available observations. As a matter of fact, we still lack a self-consistent explanation 
of the physical processes at the basis of MP formation (e.g., see \citealt{bastianlardo18,gratton19}).

Understanding the kinematical and structural properties of MPs can provide new insights into the early epochs of GC formation and evolution. 
In fact, most formation models suggest that MPs form with different structural and kinematic properties. Differences between the FP and the SP kinematics can be either imprinted at the time of SP formation (see e.g., \citealt{bekki10,lacchin22}) or emerge during a cluster’s evolution as a consequence of the initial differences between the FP and SP spatial distributions (see e.g., \citealt{tiongco19,vesperini21,sollima21}). 
Although the primordial structural and kinematic differences between FP and SP stars are expected to be gradually erased during GC long-term dynamical evolution (e.g., \citealt{vesperini13,brunet15,tiongco19,vesperini21,sollima21}), some clusters are expected to still retain some memory of these initial differences.
Indeed, \citet{dalessandro19} measured the difference in the spatial distributions of FP and SP stars for a large sample of massive Galactic and Magellanic Clouds clusters homogeneously observed with HST. The authors found that the differences between the FP and SP spatial distributions generally follow the evolutionary 
sequence expected for the long-term dynamical evolution of clusters forming with an initially more centrally concentrated SP subsystem 
(see also \citealt{leitinger23,onorato23,cadelano24}).

Spatial distributions alone can provide only a partial picture of the dynamical properties of MPs and further key constraints on 
the possible formation and dynamical paths of MPs are expected to be hidden in their kinematic properties.
Because of the technical limitations to derive kinematic information for large and significant samples of resolved stars in dense environments, most of the available information so far have been obtained by using HST proper motions (PMs) 
and ESO/VLT MUSE line-of-sight (\texttt{LOS}) velocities sampling relatively small portions of the cluster and focusing typically on the innermost regions. 
In a few particularly well studied systems, MPs have been found to show different degrees of orbital anisotropy (e.g., \citealt{richer13,bellini15,libralato23}) and possibly different rotation amplitudes 
(e.g., \citealt{cordero17,kamann20,cordoni20,dalessandro21a,martens23}). In other cases however, no significant differences between the MP kinematic properties have been observed (see for example \citealt{milone18,cordoni20,libralato19,szigeti21,martens23}) and as a consequence, the emerging internal kinematic results describe a pretty heterogeneous picture. 

To move a leap forward in our understanding of MP kinematic properties and their possible implications on GC formation, it is fundamental to perform a systematic and homogeneous study of clusters sampling a wide range of dynamical ages and including the analysis of the clusters' outer regions, which are expected to retain some memory of the primordial structural and kinematic differences for longer timescales. 
As a first step in this direction, in this paper, we perform for the first time a self-consistent study of the 3D kinematics of MPs in a representative sample of Galactic GCs for which it is possible to 
sample virtually their entire radial extension.  
This study has the additional advantage to overcome the typical limitations connected with projection effects, which typically arise when \texttt{LOS} or plane-of-the-sky velocities (i.e., PMs) are used independently possibly hampering the detection of the actual differences between the MP kinematic properties (see e.g. the discussion concerning these issues in \citealt{tiongco19}).
 
The paper is structured as follows. In Section~2 the adopted sample and the observational database are presented. 
In Section 3 we describe the kinematic analysis along the three velocity components, while in Section~4 we present the approach adopted for the study of the morphological properties of MPs.
In Sections~5 and 6 we present the main observational results along with a detailed comparison with dynamical simulations, and in Section~7 we compare them with the literature. 
In Section~8 we summarize our findings and discuss their possible implications in the context of massive clusters formation and early evolution.

\begin{table*}
\caption{Properties of the 16 clusters fully analyzed in the present work.}
\centering
\begin{tabular}{lccccccc}
\label{tab:gc}\\
\hline \hline
Cluster & D [kpc] &  $r_{h} [\arcsec]$ & Log$(t_{rh})$ & age [Gyr] &  $N_{RV}$ & $N_{PM}$ \\
\hline
NGC 104 (47 Tuc) &   4.5        &  190.2   & 9.55  & 12.75 &  1190 & 2427 \\
NGC 288          &   8.9        &  133.8   & 9.32  & 12.50 &   293 &  519  \\
NGC~1261         &  16.3        &   40.8   & 9.12  & 11.50 &    99 & 291    \\
NGC 1904 (M 79)  &  12.9        &   39.0   & 8.95  & 12.50 &   214 & 415   \\
NGC 3201         &   4.9        &  186.0   & 9.27  & 12.00 &   415 &  664     \\
NGC 5272 (M 3)   &  10.2        &  138.6   & 9.79  & 12.50 &   370 & 900      \\
NGC 5904 (M 5)   &   7.5        &  106.2   & 9.41  & 12.25 &   480 & 787      \\
NGC 5927         &   7.7        &   66.0   & 8.94  & 12.25 &   137 & 619  \\  
NGC 5986         &  10.4        &   58.8   & 9.18  & 13.25 &   160 & 633      \\
NGC 6093 (M 80)  &  10.0        &   36.6   & 8.80  & 13.50 &   433 & 668      \\
NGC 6205 (M 13)  &   7.1        &  101.4   & 9.30  & 13.00 &  313 & 1201 \\    
NGC 6362         &   7.6        & 123.0    & 9.20  & 12.50 &   489 & 713   \\
NGC 6171 (M 107) &   6.9        & 103.8    & 9.00  & 12.75 &   184 & 379   \\
NGC 6254 (M 10)  &   4.4        & 117.0    & 8.90  & 13.00 &  296 & 589   \\
NGC 6496         &  11.3        &  61.2    & 9.04  & 12.00 &   92 & 174   \\
NGC 6723         &   8.7        &  91.8    & 9.24  & 12.75 &  251 & 515  \\
\hline
\end{tabular}
\tablefoot{Distances are from \citet{baumgardt19}, structural parameters from \citet{harris96} and ages come from the compilation by \citet{dotter10} with the exception of NGC~1904 for which we used the age derivation by \citet{dalessandro13}. N$_{LOS}$ and N$_{PM}$
represent the number of \texttt{LOS} velocities and PMs used for the kinematic analysis.}
\end{table*}

\section{Sample definition and observational data-sets}
The analysis presented in this paper targets 16 Galactic GCs. 
In detail, the sample includes all clusters analyzed by \citet{ferraro18} and \citet{lanzoni18a,lanzoni18b} in the context of the {\it ESO/VLT Multi Instrument Kinematic Survey of Galactic GCs} (MIKiS) but NGC~362 as it lacks near-UV photometric data
needed for the study of MPs (see Section 2.2 for details). We added to the target list NGC~104 (47~Tucanae) as it is a massive, relatively close and well studied GC, which can be useful for comparative analysis. We included also NGC~6362 for which we secured a large kinematic data-set in \citet{dalessandro18b,dalessandro21a}, 
NGC~6089 (M~80) and NGC~6205 (M~13) as they have been found 
to show interesting kinematic properties in previous analysis by \citet{cordero17} and \citet{kamann20}. 
Table~\ref{tab:gc} summarizes some useful properties of the targets, such as distances, structural properties, age, relaxation times and kinematic sample sizes.
The selected clusters are representative of the overall Galactic GC population as they properly encompass the cluster dynamically-sensitive parameter space, spanning a large range of central
densities and concentrations, different stages of dynamical evolution and different environmental conditions. 
They are also more massive than $M>10^4$ $M_{\odot}$ and relatively close to the Earth 
(within $\sim16$ kpc), thus providing data with good signal-to-noise ratios (S/Ns) for a large sample of stars. 

\begin{figure}
\centering
\includegraphics[width=\columnwidth]{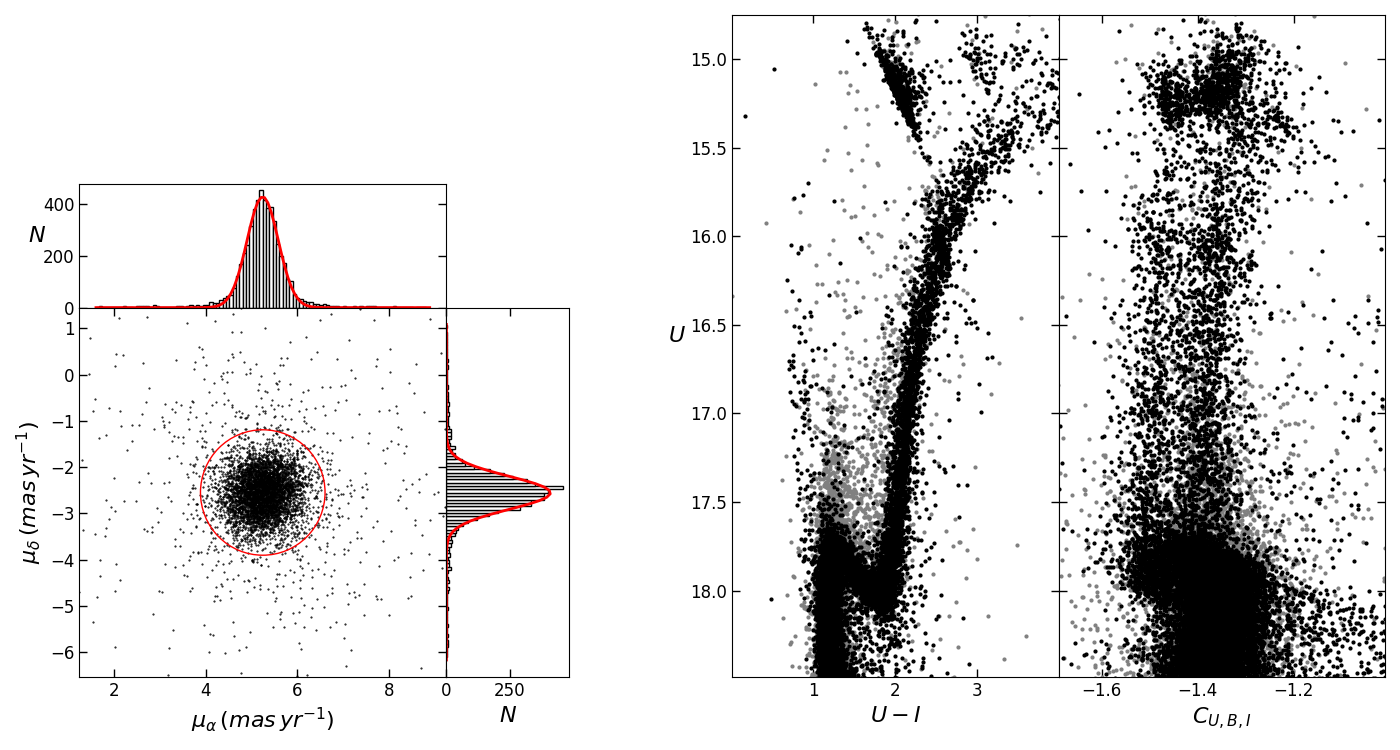}
\caption{On the left, the vector-point-diagram obtained by using {\it Gaia} DR3 PMs for the GC 47 Tuc is shown along with the distribution of stars along the $\mu_{\alpha}^*$ and $\mu_{\delta}$ velocity components. The red circle represents the $2\sigma$ selection 
described in Section~2.3.
The panel on the right shows the (U, U-I) and (U, C$_{U,B.I}$) CMDs for 47 Tuc obtained by using ground-based photometric catalogs published by \citet{stetson19}. Likely member stars based on the PM selection shown in the left panel are highlighted in black, while in grey likely field interlopers are shown.}
\label{fig:cmdpm} 
\end{figure}

\subsection{Kinematic database}
The analysis performed in this paper is based on two main kinematic data-sets securing \texttt{LOS} velocities (RVs) and PMs for hundreds (or thousands in a few cases - Table~\ref{tab:gc}) of red giant branch stars (RGBs) in each GC.
For 12 out of 16 GCs, most of the adopted \texttt{LOS} RVs were obtained by using ESO/VLT KMOS and FLAMES data obtained as part of the MIKiS survey.
We refer the reader to \citet{ferraro18} for details about the overall observational strategy and data-analysis.
MIKiS \texttt{LOS} RVs were then complemented with those from the publicly available catalog by \citet{baumgardt19} to improve the sampling of the external regions of the target clusters. All \texttt{LOS} RVs used for 47~Tuc come from the \citet{baumgardt19} catalog.
For NGC~6362, M~80 and M~13 \texttt{LOS} RVs were obtained by using ESO/VLT MUSE and FLAMES data \citep{cordero17,dalessandro18b,dalessandro21a,kamann20}.

For each of the investigated GCs, astrometric information, namely absolute PMs ($\mu_{\alpha}^*$, $\mu_{\delta}$) and relative errors, 
were retrieved from the ESA/{\it Gaia} DR3 \citep{GaiaDR3} archive
out to the clusters' tidal radius. Only stars with $\texttt{ruwe}<1.3$\footnote{\texttt{ruwe} is the {\it Gaia} Renormalised Unit Weight Error and provides a measures of the quality of the astrometric observations fit.} were then used for the kinematic analysis.

\begin{figure}
\centering
\includegraphics[width=\columnwidth]{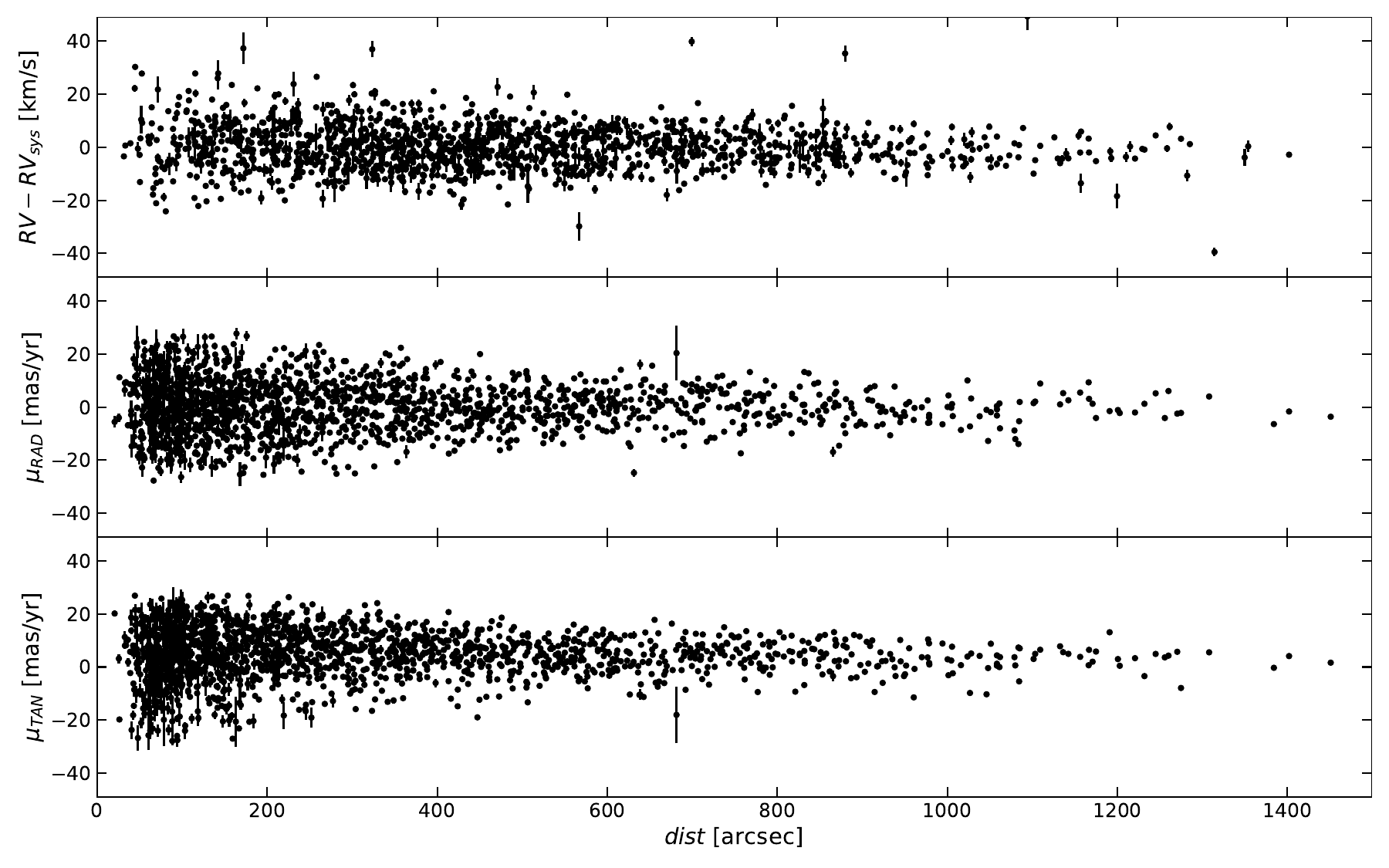}
\caption{\texttt{LOS}, \texttt{RAD} and \texttt{TAN} velocity distributions of likely member stars of the GC 47 Tuc as a function of the cluster-centric distance. All velocities are shown with respect to the cluster systemic velocity along the corresponding component.}
\label{fig:velocities}
\end{figure}

\subsection{Photometric dataset, membership selection and differential reddening correction}
We used the wide-field catalogs published by \citet{stetson19} and
including U, B, V, R and I bands to identify MPs in the target GCs (see Section~2.3) While these data are seeing-limited and can suffer of incompleteness in the crowded central regions, they have similar spatial resolution as the kinematic \texttt{LOS} RVs dataset. In addition, they are typically not affected by saturation problems and therefore they maximize the number of bright stars in common with the kinematic samples. 
The photometric catalogs were cross-matched with the kinematic ones based on their absolute coordinates ($\alpha$, $\delta$) and by using the cross-correlation tool \texttt{CataXcorr}\footnote{\texttt{CataXcorr} is a code aimed at cross-
correlating catalogs and finding geometrical transformation solutions - \url{http://davide2.bo.astro.it/~paolo/Main/CataPack.html}. 
It was developed by P. Montegriffo at INAF-OAS Bologna and it has been used by our group for more than 20 years now.}.
For each cluster in the sample, the final catalog includes all stars in common between {\it Gaia} and the photometric catalogs. 
A fraction (typically larger than $\sim60\%$ along the RGB) of these stars has also \texttt{LOS} RVs and is therefore suited for a full 3D analysis (see Table~\ref{tab:gc}). 

To separate likely cluster members from field interlopers, we selected stars whose PMs are within $2\sigma$ from the cluster systemic velocity (adopted 
from \citealt{vasiliev21}), where $\sigma$ is the standard deviation of the observed ($\mu_{\alpha}^*$, $\mu_{\delta}$) distribution of RGB stars.  We have verified that for the clusters in our sample, reasonable variations of the adopted cluster membership selection criteria do not have a significant impact on the main results of the kinematic analysis. 
As an example, Figure~\ref{fig:cmdpm} shows the PM distribution of 47~Tuc stars along with the (U, U-I) and (U, $C_{U,B,I}$ -- where $C_{U,B,I}=(U-B)-(B-I)$; \citealt{monelli13}) CMDs for both selected cluster stars and field interlopers. 
Figure~\ref{fig:velocities} shows instead the distributions of the velocities along the \texttt{LOS}, the PM 
radial (\texttt{RAD}) and PM tangential (\texttt{TAN}) components as a function of the cluster-centric distance for likely members RGB stars of the same cluster. 
It is worth mentioning here that the kinematic catalogs obtained from the MIKiS survey already rely on the cluster membership selection performed by \citet{ferraro18} and \citet{lanzoni18a,lanzoni18b}, which is  based on both the \texttt{LOS} RV and [Fe/H] distributions (we refer to those papers for further details).

Available magnitudes were then corrected for differential reddening by using the approach described in \citet{dalessandro18c} \citep[see also][]{cadelano20}.
In short, differential reddening was estimated by using likely member stars selected in a magnitude range typically going from the RGB-bump level down to about one magnitude below the cluster turn-off. By using these stars a mean ridge-line (MRL) was defined in the (B, B-I) CMD.
Then for all stars within $3\sigma$ (where $\sigma$ is the color spread around the MRL) the geometric distance from the MRL ($\Delta X$) was computed. For each star in the catalog, differential reddening was then obtained as the mean of the $\Delta X$ values of the 30 nearest (in space) selected stars. $\Delta X$ was then transformed into differential reddening $\delta E(B-V)$ by using  Eq. 1 from \citet{dalessandro18c}, properly modified to account for the different extinction coefficients for the adopted filters. 
Differential reddening corrections turn out to be relatively small ($<0.1$mag) for most GCs in the sample, with the most critical cases being NGC~3201 and NGC~5927 for which we find $\delta E(B-V)$ values larger than $\sim0.2$ mag.

\begin{figure}
\centering
\includegraphics[width=\columnwidth]{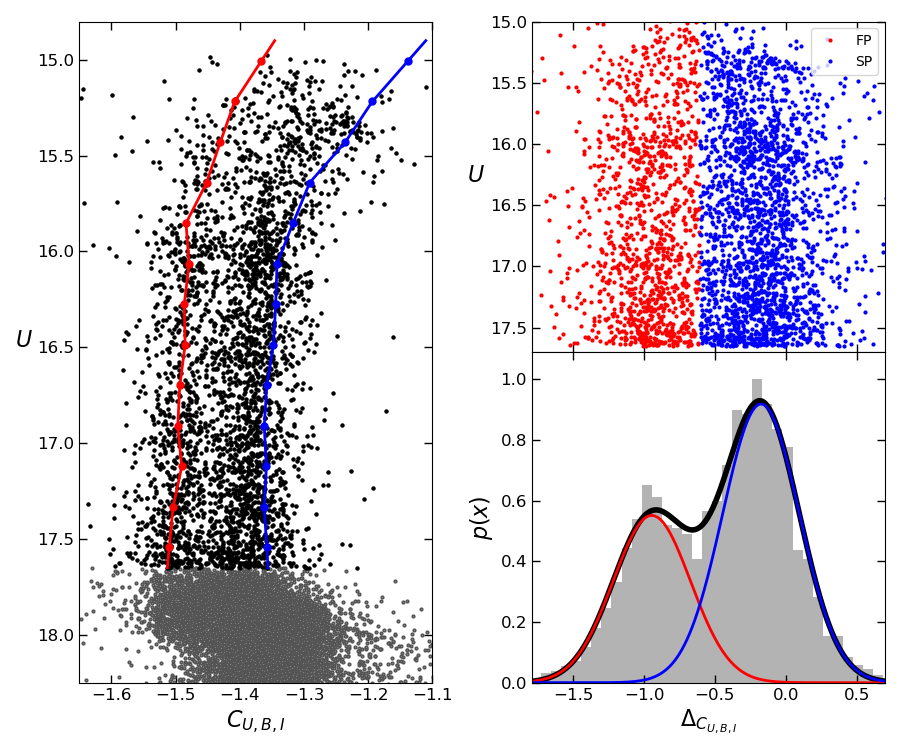}
\caption{Left panel: (U, C$_{U,B,I}$) CMD of 47 Tuc likely member stars. RGB stars adopted for the kinematic analysis are highlighted in black. The blue and red curves are the fiducial lines adopted to verticalize the color distribution. 
Right-hand panels: the top panel displays the verticalized color distribution of RGB stars, while the bottom panel shows the corresponding histograms. The red and blue
curves represent the two best-fit Gaussians for the FP and SP, respectively, while the solid black curve is their sum.}
\label{fig:cmdsel} 
\end{figure}

\subsection{MP classification}
Starting from the sample of likely member stars and differential reddening corrected magnitudes, we identified MPs along the RGB by using their distribution in the ($U$, $C_{U,B,I}$) CMD (Figure~\ref{fig:cmdsel}).  It has been shown that this color combination is very effective to identify MPs along the RGB with different C and N (and possibly He) abundances \citep{sbordone11,monelli13}. RGB stars were verticalized in the ($U$, $C_{U,B,I}$) CMD with respect to two fiducial lines on the blue and red edges of the RGB, calculated as the 5th and 95th percentile of the color distribution in different magnitude bins \citep[Figure~\ref{fig:cmdsel} -- see, e.g.,][for similar implementations of the same technique]{dalessandro18a,dalessandro18c,onorato23,cadelano23}. In the resulting verticalized color distribution ($\Delta_{C_{U,B,I}}$; right panel in Figure~\ref{fig:cmdsel}), 
stars on the red (blue) side are expected to be N-poor (-rich), i.e., FP (SP), respectively. 
We run a two components Gaussian Mixture Modeling\footnote{We use the 
scikit-learn package \citep{pedregosa11}.} (GMM) analysis to the resulting $\Delta_{C_{U,B,I}}$
distribution thus assigning to each star a probability of belonging to the FP and SP sub-populations. Stars with a probability larger 
than $50\%$ to belong to one of the two sub-populations were then flagged as FPs or SPs. 
Figure~\ref{fig:cmdsel} shows the result of the MP identification and separation for the GC 47~Tuc.
While this approach may introduce a few uncertainties and over-simplifications in the MP classification as we are not directly deriving light-element chemical abundances, it secures statistically large samples of stars with MP tagging that are hard to obtain by using only spectroscopic data.

%%%%%%%%%%%%%%%%%%%%%%%%%%%%%%%%%%%%%%%%%%%%
\section{Kinematic analysis}
For each cluster in the sample we first analyzed the kinematic properties in terms of velocity dispersion, rotation and anisotropy profiles for the \texttt{LOS} and plane-of-the-sky components separately (Sections 3.1 and 3.2) and then, for the fraction of stars for which all velocity components are available, we performed a full 3D study (Section~3.3).  
We adopted as clusters' centers those reported by \citet{goldsbury10}.
All velocities were corrected for perspective effects induced by the clusters' systemic motions by using the equations reported in \citet{vanleeuwen09} and following the approach already adopted in \citet{dalessandro21b,dellacroce23}.

\begin{figure*}
\centering
\includegraphics[width=\textwidth]{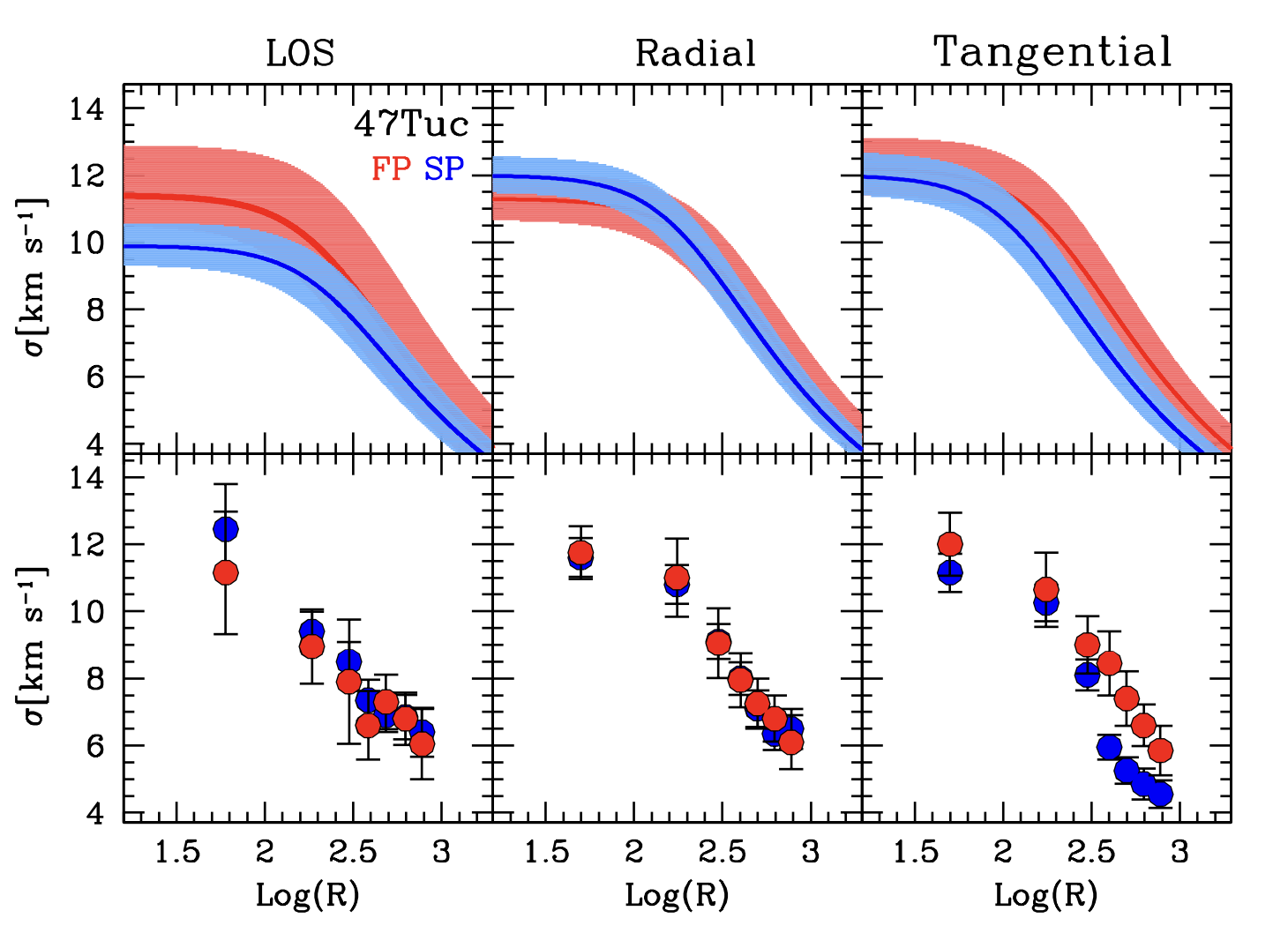}
\caption{{\it Bottom panels}: observed velocity dispersion profiles along the \texttt{LOS}, radial and tangential velocity components for MPs in 47 Tuc obtained by using a maximum-likelihood approach on binned data.
{\it Upper panels:} Best-fit velocity dispersion profiles as obtained by using the bayesian analysis on discrete velocities described in Section 3.1.}
\label{fig:47Tuc_sig}  
\end{figure*}

\subsection{1D velocity dispersion and rotation profiles}
To characterize the kinematic properties of the clusters in the sample and of their sub-populations, we adopted the Bayesian approach described in \citet{cordero17} (see also \citealt{dalessandro18b}), which is based on the use of a discrete fitting technique 
to compare simple kinematic models (including a radial dependence of the rotational amplitude and velocity dispersion of the cluster) with individual radial velocities.
We stress that this is a purely kinematic approach aimed at searching for relative differences among different clusters and sub-populations and it is not aimed at providing a self-consistent dynamical description of each system.

The likelihood function for the radial velocities of individual stars depends on our assumptions about the formal descriptions of the rotation and velocity dispersion radial variations.
For the velocity dispersion profile we assumed the functional form of the Plummer model \citep{plummer11}, which is simply defined by its central velocity dispersion $\sigma_0$ and its scale radius $a$:

\begin{equation}
\label{sigma2R}
\sigma^2(R) = \frac{\sigma_0^2}{\sqrt{1 + R^2/a^2}} \ ,
\end{equation}

where $R$ is the projected distance from the centre of the cluster. We adopted the same formal description for all velocity components.
For the rotation curve, we assumed cylindrical rotation and adopted the functional form expected for stellar systems undergoing violent relaxation during their early phases of evolution \citep{lyndenbell67}:

\begin{equation}
\label{rotation_los}
V_{\rm rot} {\rm sin}i (X_{\rm PA_0}) = \frac{2A_{\rm rot}}{R_{\rm peak}} \frac{X_{\rm PA_0}}{1+(X_{\rm PA_0}/R_{\rm peak})^2}
\end{equation}

\begin{equation}
\label{rotation_pm}
\mu_{\rm TAN} = \frac{2V_{\rm peak}}{R_{\rm peak}} \frac{R}{1+(R/R_{\rm peak})^2}
\end{equation}

for the \texttt{LOS} (Eq.~2) and \texttt{TAN} (Eq.~3) velocity components, respectively. In Eq.~2 V$_{\rm rot} {\rm sin}i$ represents the projection of the rotational amplitude along the \texttt{LOS} velocity component at a projected distance $X_{\rm PA_0}$ from the rotation axis. $A_{\rm rot}$ is the peak rotational amplitude occurring at the projected distance $R_{\rm peak}$ from the cluster center. 
We defined the rotation axis position angle ($PA$) as increasing anti-clockwise in the plane of the sky from north (PA=$0^{\circ}$) towards east (PA=$90^{\circ}$).
Since the inclination of the rotation axis is unknown, V$_{\rm rot} {\rm sin}i$ represents a lower limit to the actual rotational amplitude. As an extreme case, if the rotation axis is aligned with the line of sight, the rotation would be in the plane-of-the-sky.
In Eq.~3, V$_{\rm peak}$ represents the maximum (in an absolute sense) of the mean motion in the \texttt{TAN} component.

\begin{figure*}
\includegraphics[width=\textwidth]{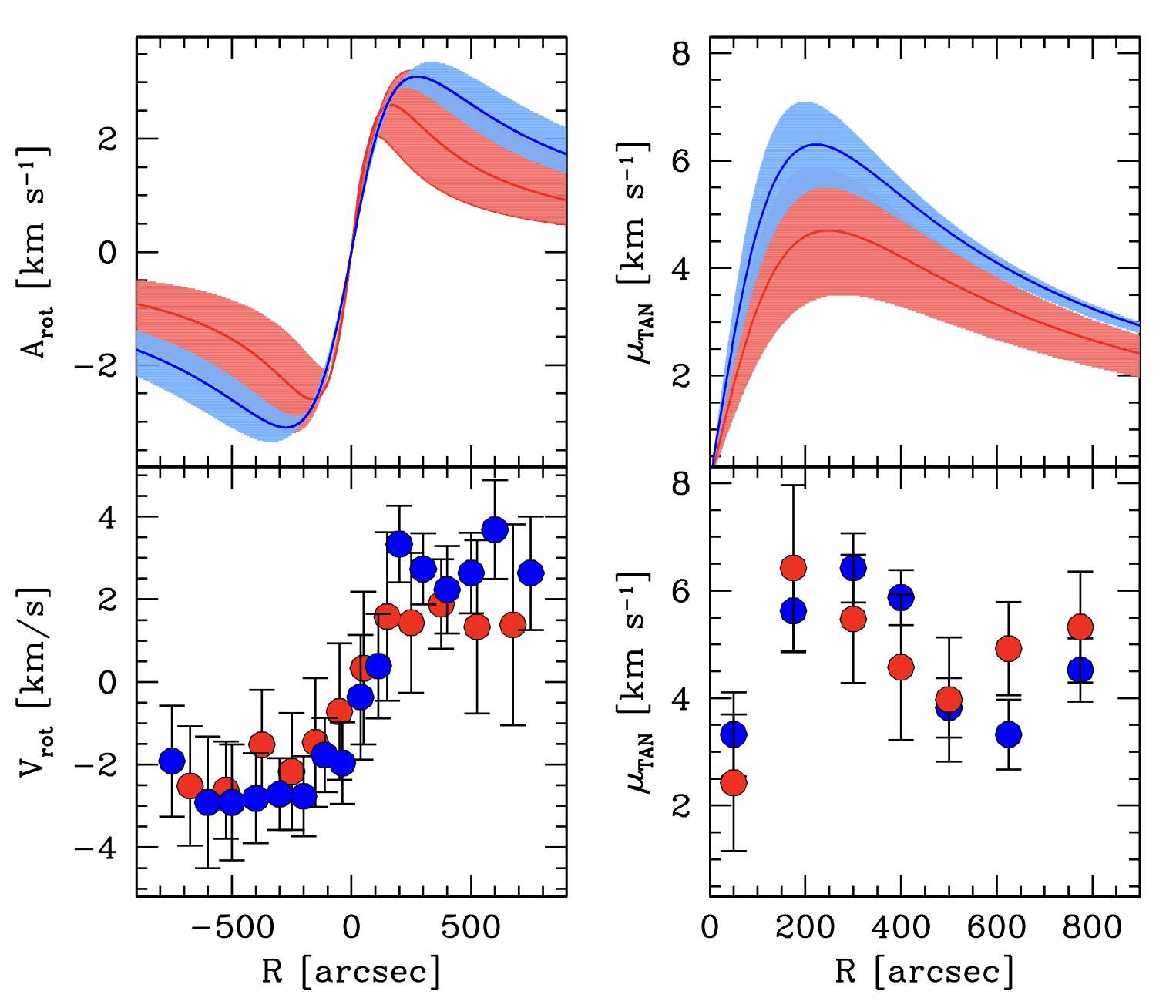}
\caption{Same as \ref{fig:47Tuc_sig}, but now for the rotation profiles along the LOS (left panel) and tangential (right panel) components for MPs in 47~Tuc.}
\label{fig:47Tuc_rot}  
\end{figure*}

The fit of the kinematic quantities was performed by using the \texttt{emcee}\footnote{\url{https://emcee.readthedocs.io/en/stable/}} (\citealt{foremanmackey13}) implementation 
of the Markov chain Monte Carlo (MCMC) sampler, which provides the posterior probability distribution function (PDFs) 
for $\sigma_0$, $a$, $A_{\rm rot}$ and $PA_0$. For each quantity, the 50th-, 16th- and 84th-percentile of the PDF distributions were adopted as the best-fit values and relative errors, respectively. 
We assumed a Gaussian likelihood and flat priors on each of the investigated parameter within reasonably large range of values.
It is important to note that in general, since the analysis is based on the conditional probability of a velocity measurement, given the position of a star, 
our fitting procedure is not biased by the spatial sampling of the stars in the different clusters and sub-samples. However, the kinematic properties are better constrained in regions that are better sampled (i.e. larger number of stars with available kinematic information).

For sanity check and comparison, we also derived the velocity dispersion and rotation profiles by splitting the surveyed areas in a set of concentric annuli, whose width was chosen as a compromise between a good radial sampling and a statistically significant number of stars. In this case the analysis was limited radially within a maximum distance from the center of the clusters to guarantee a symmetric coverage of the field of view. The adopted limiting distance varies from one cluster to the other depending on the photometric and kinematic dataset field of view limits (Section~2.3).
While this approach requires the splitting of the sample in concentric radial bins, whose number and width is at least partially arbitrary and can potentially have an impact on the final results, it has the advantage of avoiding any assumption on the model description of the velocity dispersion and rotation profiles. 

In each radial bin the velocity dispersion was computed by following the maximum-likelihood approach described by \citet{pryor93}. 
The method is based on the assumption that the probability 
of finding a star with a velocity of $v_i$ and error $\epsilon_i$ at a projected distance from the cluster center $R_i$ can be approximated as:

\begin{equation}
    p(v_i,\epsilon_i,R_i) = \frac{1}{2\pi\sqrt{\sigma^2 + \epsilon_i^2}}exp{\frac{(v_i-v_0)^2}{-2(\sigma^2 + \epsilon_i^2)}}
\end{equation}

where $v_0$ and $\sigma$ are the systemic velocity and the intrinsic dispersion profile of the cluster along the three components (i.e., \texttt{LOS}, \texttt{RAD} and \texttt{TAN}), respectively.

As for the rotation along the \texttt{LOS} component, we used the method fully described in \citet{bellazzini12} and adopted in the previous papers of our group \citep[e.g.,][]{ferraro18,lanzoni18a,dalessandro21a,leanza22}. In brief, we considered a line passing through the cluster center with position angle varying from $-90^{\circ}$ to $90^{\circ}$ by steps of $10^{\circ}$. For each value of $PA$, such a line splits the observed sample in two.  If the cluster is rotating along the line of sight, we expect to find a value of $PA$ that maximizes the difference between the median RVs of the two sub-samples, since one component is mostly approaching and the other is receding
with respect to the observer.
Moving $PA$ from this value has the effect of gradually decreasing the difference in the median RV.
Hence, the appearance of a coherent sinusoidal behavior as a function of $PA$ is a signature of rotation and its best-fit sine function provides an estimate of the rotation amplitude ($A_{\rm rot}$) and the position angle of the cluster rotation axis (PA$_0$). For the plane-of-the-sky rotation we used instead the variation of the mean values within each radial bin of the tangential velocity component with respect to the systematic motion. 

Examples of the results obtained with both the Bayesian and maximum-likelihood analyses are shown in Figures~\ref{fig:47Tuc_sig} and ~\ref{fig:47Tuc_rot} for the MPs of the GC 47~Tuc.
For all clusters in the sample we find a good agreement between the discrete and binned analysis, however in the following 
we will adopt the best-fit results (and errors) obtained with the Bayesian approach. 
Table~\ref{tab:mp_kin} reports the best-fit values and relative errors for the most relevant quantities
along both the LOS and plane-of-the-sky for both the FP and SP.

%%%%%%%%%%%%%%%%%%%%M
\subsection{Anisotropy profiles}

\begin{figure*}
\centering
\includegraphics[width=0.95\textwidth]{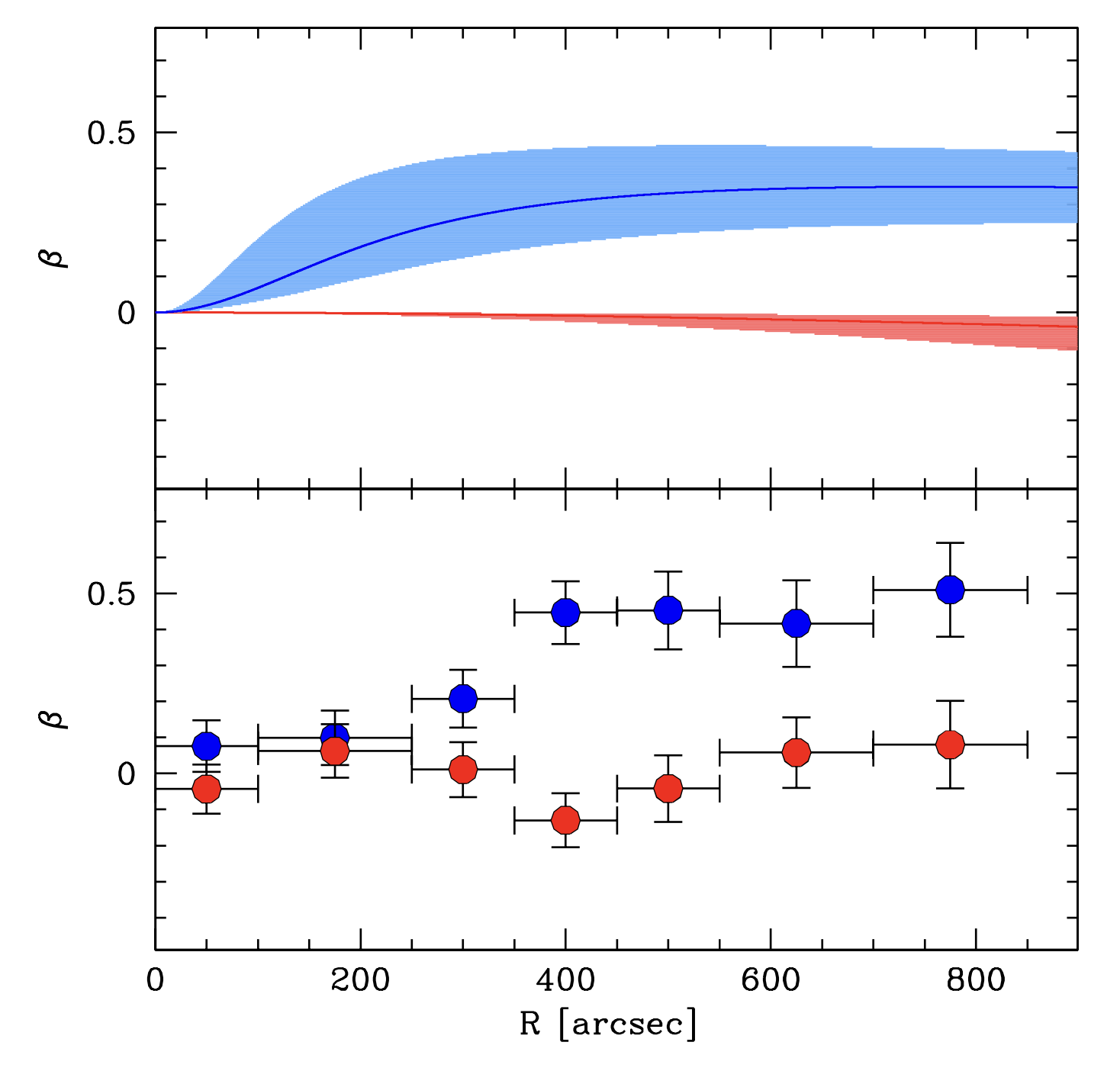}
\caption{Same as \ref{fig:47Tuc_sig}, but now for the anisotropy profile (see Section 3.2).}
\label{fig:47Tuc_ani}  
\end{figure*}
 
We adopted a modified version of the Osipkov-Merrit \citep{osipkov79,merritt85} 
model to provide a formal description of the velocity anisotropy profile for each cluster and sub-population 
(see Aros et al., in prep. for further details).
This model is
isotropic in the cluster centre and it becomes increasingly anisotropic for $R>r_{a}$ (where $r_{a}$ is the anisotropy radius). 
Our anisotropy description includes two additional parameters, which are the `outer' anisotropy value ($\beta_{\infty}$) and the truncation radius ($r_{t}$):

\begin{equation}\label{ani}
\beta(R) = \frac{\beta_{\infty} R^2}{(R^2 + r_{a})^2}\left(1-\frac{R}{r_{t}}\right)
\end{equation}

$\beta_{\infty}$ defines how radial or tangentially biased the velocity anisotropy profile is for $R>>r_{a}$, and $r_{t}$ is the radius at which the velocity anisotropy becomes isotropic again. In practice, we adopt $r_{t}$ as the Jacobi radius of the cluster.

Following the same approach adopted for the velocity dispersion and rotation, we performed a MCMC fit of the anisotropy profiles (see results in Table~\ref{tab:mp_kin}) 
by using the individual stellar velocities along the radial and tangential components ($\mu_{\rm RAD}$ and $\mu_{\rm TAN}$, respectively) and by assuming a Gaussian likelihood in which 
the best-fit velocity dispersion along the radial direction $\sigma_{\rm RAD}$ can be described through a Plummer profile 
(see Equation~\ref{sigma2R}) and the velocity dispersion along the tangential direction is simply given by $\sigma^2_{\rm TAN}=[1-\beta(R)]\sigma^2_{\rm RAD}$. 
It is important to stress that this assumption is only adopted for the best-fit anisotropy parameters derivation, while we will generally consider as best-fit velocity dispersion profiles those derived through the independent fit of $\sigma_{\rm RAD}$ and $\sigma_{\rm TAN}$ discussed in Section~3.1. 
For the binned analysis, the anisotropy profiles were obtained directly from the \texttt{RAD} and \texttt{TAN} velocity dispersion values obtained in Section~3.1.
Figure~\ref{fig:47Tuc_ani} shows an example of the best-fit anisotropy profile for the MPs of 47~Tuc.

\subsection{Full 3D analysis}
To perform a full 3D analysis we used the kinematic sample of member stars having both \texttt{LOS} RVs and {\it Gaia} PMs after quality selection (see Section~2.1).
We also limited the analysis to the same radial extension adopted for the binned maximum-likelihood analysis (Section~3.1).

We followed the approach described in \citet{sollima19}, which has the advantage of constraining a cluster full rotation pattern by estimating the inclination angle of the rotation axis ($i$) with respect to the line of sight, the position angle of the rotation axis ($\theta_0$) and the rotation velocity amplitude ($A$), by means of a model-independent analysis. We refer the reader to that paper for a full description of the method. Here we only report the main ingredients and assumptions. 

In a real cluster the angular velocity is expected to be a function of the distance from the rotation axis (see for example Eqs.~1 and ~2). 
To account for such a dependence in a rigorous way, a rotating model should be fitted to the data. However, 
to perform a model-independent analysis we considered an average projected rotation velocity with amplitude $A_{\rm 3D}$, which has been assumed to be independent on the distance from the cluster center. 
While of course, this represents a crude approximation of the rotation patterns expected in GCs and provides a rough average of the actual rotation amplitude,
it is important to stress that it does not introduce any bias in the estimate of $\theta_0$ and $i$.

\begin{figure*}
\centering
\includegraphics[width=0.95\textwidth]{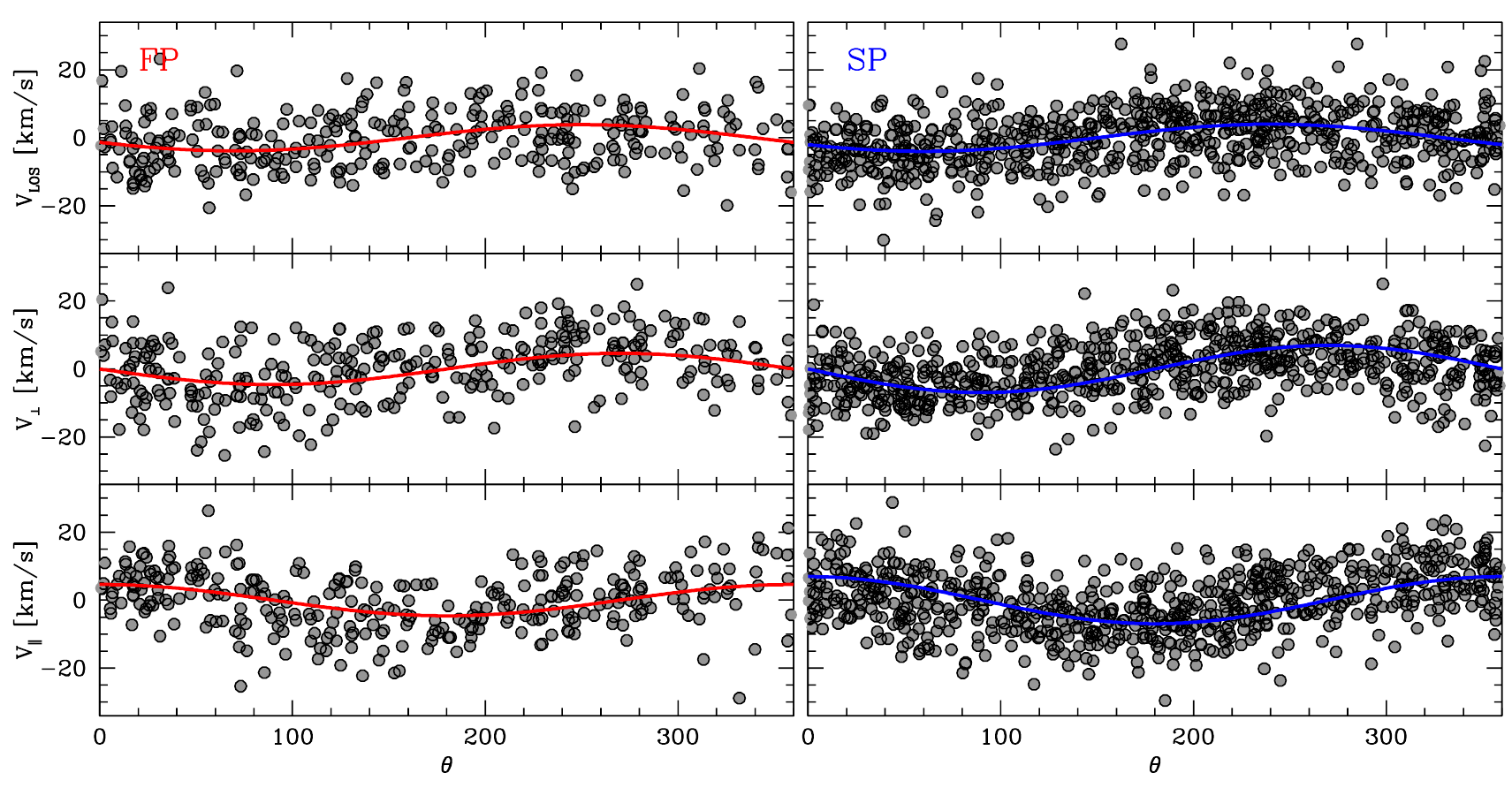}
\caption{Distribution of the three velocity components as a function of the position angle for MPs in 47 Tuc.
Left and right panels refer to the FP and SP sub-population, respectively. The solid lines show the best-fitting trend in all panels.}
\label{fig:47Tuc_3D}
\end{figure*}

$A_{\rm 3D}$, $i$ and $\theta_0$ were derived by solving the equations describing the rotation projection along the \texttt{LOS} ($V_{\rm LOS}$) and those perpendicular ($V_{\perp}$) and parallel ($V_{\parallel}$) to the rotation axes (see Eq. 2 in \citealt{sollima19}). 
While the velocity component perpendicular to the rotation axis has a dependence on stellar positions within the cluster along the line of sight, we neglected it in the present analysis. In fact, we note that the dependence on the stellar distance along the line of sight does not affect the mean trend of the perpendicular velocity component, but it can only introduce an additional spread on its distribution.  
We assumed that $i$ varies in the range $0^{\circ}<i<90^{\circ}$ with respect to the line of sight and
the position angle in the range $0^{\circ}<\theta_0<360^{\circ}$. $\theta_0$ grows anti-clockwise from North to East 
and $A_{\rm 3D}$ is positive for clockwise rotation in the plane of the sky.
Following the approach already adopted for the 1D analysis, we derived the best-fit rotation amplitudes, 
position and inclination angles and relative errors by maximizing the 
likelihood function reported in Eq. 3 of \citet{sollima19} by using the MCMC algorithm \texttt{emcee}.
Best-fit results are reported in Table~\ref{tab:mp_kin}. 
Figure~\ref{fig:47Tuc_3D} shows the result of the best-fit analysis along the three velocity components for the FP and SP sub-populations of 47~Tuc.

\begin{figure*}
\centering
\includegraphics[width=0.95\textwidth]{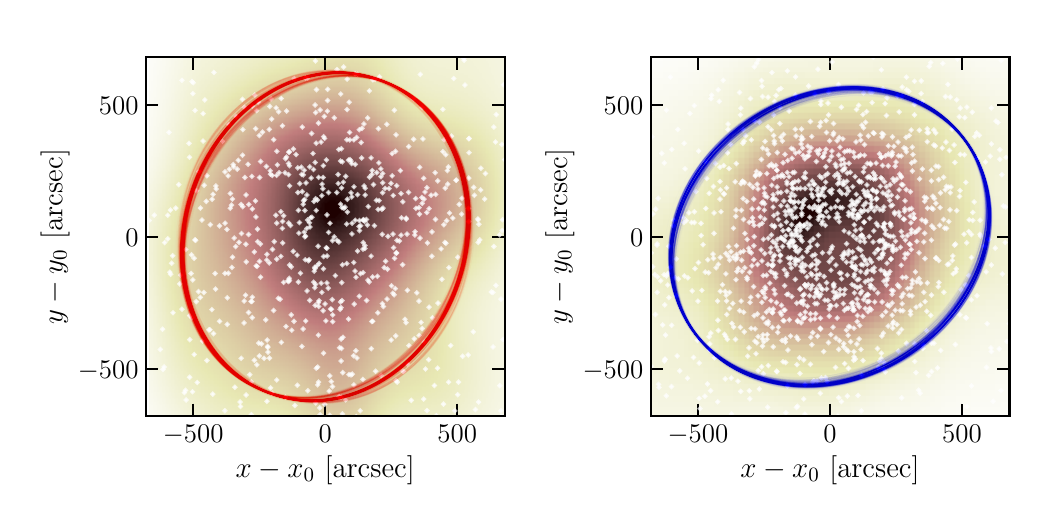}
\caption{2D density maps of the stars selected in the GC 47 Tuc for the kinematic analysis. 
FPs are shown on the left panel, while SPs in the right panel. Overplotted to the density distributions 
are the best-fit ellipses derived as described in Section~4.}
\label{fig:ell}
\end{figure*}

\section{Morphological analysis: MP ellipticity}
We inferred the ellipticity of the MPs by constructing the so-called 
shape tensor \citep{zemp11}, which is defined by the following equation
\begin{equation}
    S_{\rm ij} \equiv \frac{\sum_{\rm k} m_{\rm k} (R_{\rm k})_{\rm i} (R_{\rm k})_{\rm j}}{\sum_{\rm k} m_{\rm k}}\,,
    \label{eq:shape-tensor}
\end{equation}
where $R_{\rm k}$ and $m_{\rm k}$ are the projected distance from the cluster center and the mass of the $k$-th star, 
within the $i$-th and $j$-th element of the shape tensor grid. 
For the purpose of this study, we assumed that all stars have the same mass.
This assumption is well justified as the adopted sample consists of RGB stars (the same used for the kinematic analysis), whose mass variations for a given cluster is expected to be $\sim0.02M_{\odot}$, depending on the metallicity.

The shape tensor was computed starting from a spherical grid, whose nodes are the stellar radial distribution's 10th, 50th, and 90th percentiles.
We adopted an iterative procedure for each bin during which the shape tensor is initially constructed from spherical distances.  
After that, being $(w_0; w_1)$ the eigenvalues (with $w_0>w_1$) and $(\boldsymbol{v}_0; \boldsymbol{v}_1)$ the respective eigenvectors of the shape tensor, it follows \citep{zemp11} that

\begin{equation}
    \epsilon = 1-\frac{w_1}{w_0}\quad{\rm and}\quad {\rm PA} = \arctan \frac{v_{\rm 0,y}}{v_{\rm 0,x}}\,.
\end{equation}

The particle coordinates were then rotated by the angle $-$PA, and distances to the center were defined by means of the circularized distance
\begin{equation}
    R_{\rm ell} \equiv \sqrt{x{'}^2 + \frac{y{'}^2}{\epsilon^2}}\,,
\end{equation}
where $(x{'}; y{'})$ are the rotated locally-Cartesian coordinates of the stars.

Finally, stars were binned in the new coordinate system according to the initial grid and the shape tensor has been computed using
$R_{\rm ell}$ instead of $R$.
Such procedure was then iterated until a relative precision of $5\%$ on the axis ratio was reached. 
Best-fit ellipticity values and relative errors have been obtained by means of a bootstrapping analysis. 
In detail, the shape tensor was computed 1000 times for each cluster by randomly selecting at each time a sub-sample 
including only $90\%$ of stars. The values corresponding to the 50th percentile of the distribution of all the 
$\epsilon$ values has been then adopted as best-fit, while the errors correspond to the 16th and 84th percentiles.  
As an example, Figure~\ref{fig:ell} shows the result of the analysis for the MPs in 47~Tuc, while 
Table~\ref{tab:mp_kin} reports the best-fit values for each GC and sub-population.

\section{Results: rotation}
To obtain quantitative and homogeneous estimates of the possible kinematic differences among MPs, to follow their evolution and eventually to compare the results obtained for all GCs in the sample with theoretical models and dynamical simulations, 
we introduced a few simple parameters described in detail in the following. These parameters are meant to incorporate in a meaningful way all the main relevant physical quantities at play in a single value.
The general approach of our analysis is not to focus on the detection of specific and particularly significant kinematic differences of specific targets, but rather to compare the general kinematic behaviors described by MPs in all targets in the sample in the most effective way.  

A description about the approach adopted to quantify the possible impact of the intrinsically limited statistical 
kinematic samples and of their incompleteness on the final results is discussed in Appendix~\ref{Appendix B}. 
Here we briefly stress that the main effects are not on the derived best-fit values, but rather on their uncertainties.
While the main focus of the following sections is on the MP kinematics, we analyzed also the entire sample 
of stars with kinematic information (hereafter labeled as TOT) for comparison with previous works and we 
present the main results in Appendix~\ref{Appendix C}.

\begin{figure*}
\centering
\includegraphics[width=0.95\textwidth]{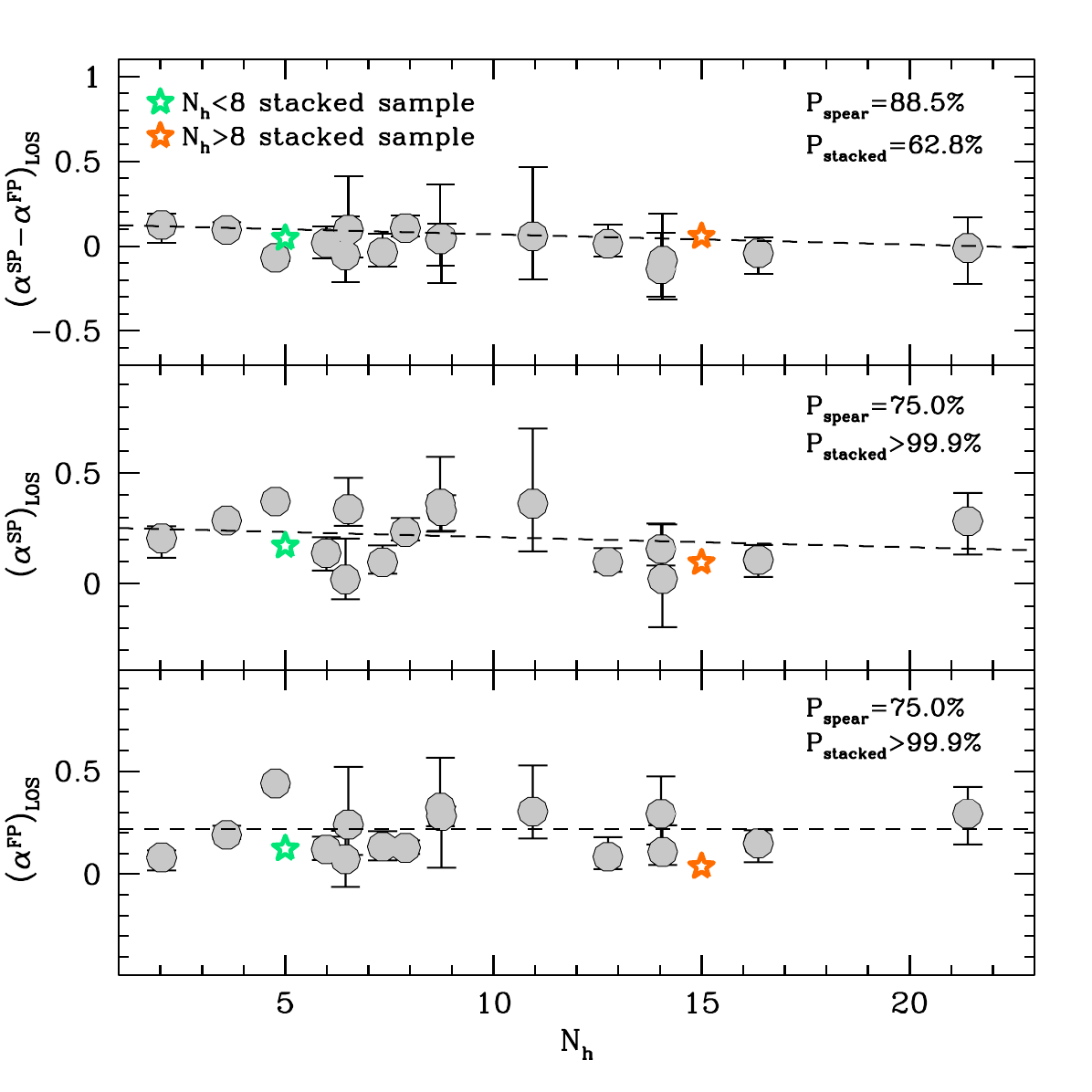}
\caption{Bottom and middle panels show the distribution of the ($\alpha)_{\rm LOS}$ parameter for the FP and SP as a function of the dynamical age $N_{h}$ for all clusters in the sample (grey circles). The upper panel shows instead the distribution of the rotation differences ($\alpha^{\rm SP} - \alpha^{\rm SP})_{\rm LOS}$. 
The dashed-lines represent the linear best-fit to the GC distribution.
In all panels, the starry symbols refer to the results obtained for the stacked analysis on the dynamically young (green) and old (orange) samples. The size of the star matches the amplitude of the errorbars.
 }
\label{fig:rot_mp}
\end{figure*}

\begin{figure*}
\centering
\includegraphics[width=0.95\textwidth]{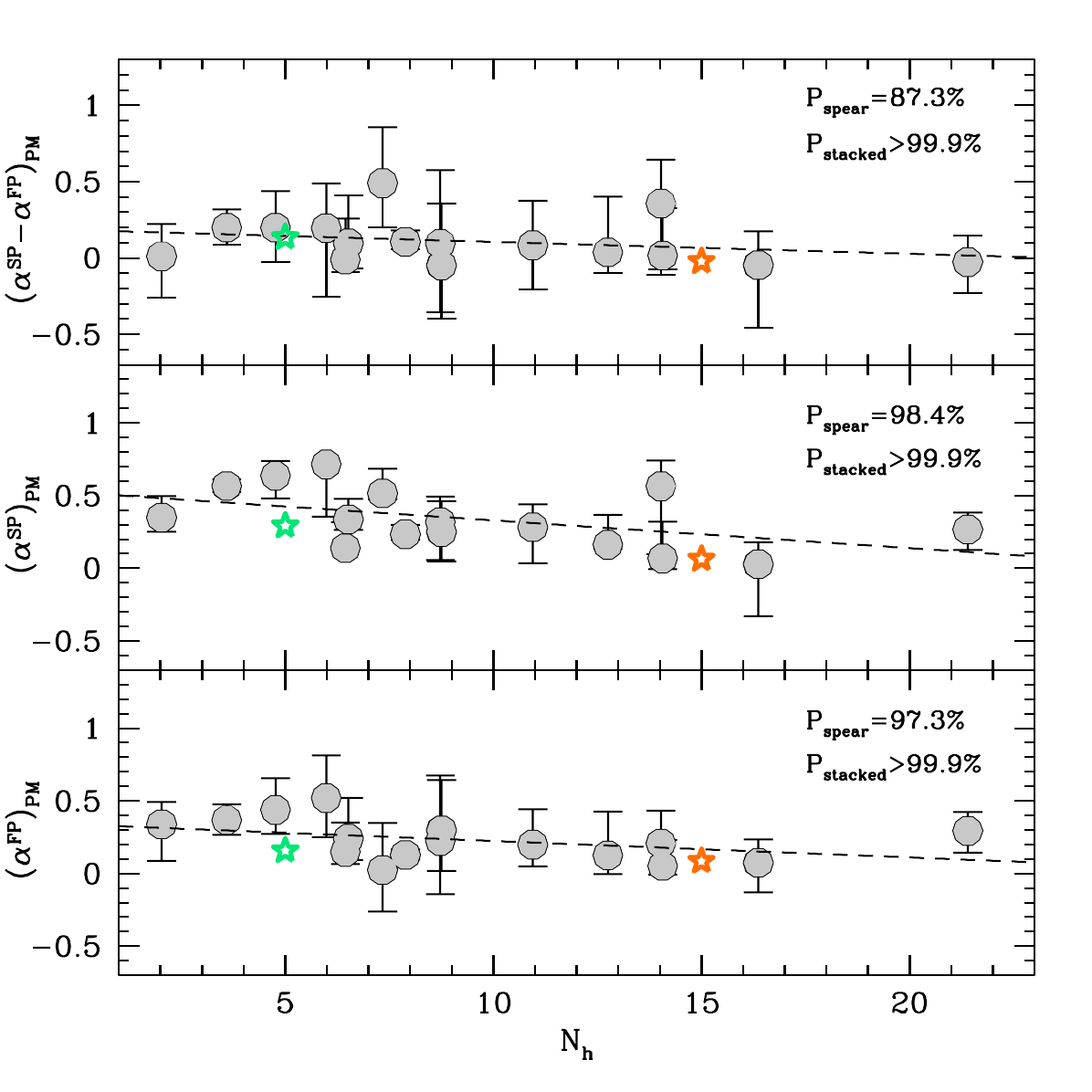}
\caption{Same as in Figure~\ref{fig:rot_mp}, but now for ($\alpha)_{PM}$ (Section~5.1).}
\label{fig:rot_mp_pm}
\end{figure*}

\subsection{Observations}
To measure the rotation differences between the SP and FP sub-populations for all clusters in the sample 
for both the \texttt{LOS} and \texttt{TAN} velocity components, we introduced a parameter, 
hereafter referred as $\alpha$, defined as the area subtended by the ratio between the best-fit rotation velocity profile and the best-fit velocity dispersion profile for each sub-population in a cluster (Section 3.1) after rescaling the cluster-centric distance to the value of the peak ($R_{\alpha}$) of such a distribution:

\begin{equation}
\alpha^{X}=\int_0^{1} V_{\rm rot}(R_{\rm l})/\sigma(R_{\rm l}) dR_{\rm l}
\end{equation}
where $V_{\rm rot}(R_{\rm l})$ and $\sigma(R_{\rm l})$ can either refer to the \texttt{LOS} or the \texttt{TAN} velocity components,  $R_{\rm l}$ is the cluster-centric distance normalized to the $R_{\alpha}$ and the index $X$ refers to the different sub-populations (i.e., FP or SP).

This parameter has the advantage of providing a robust measure of the relative strength of the rotation signal 
over the disordered motion at any radial range without making any assumption about the underlying star or mass distribution. 
By construction $\alpha$ depends on the considered cluster-centric distance and therefore a meaningful cluster-to-cluster comparison requires that the parameter is measured over equivalent radial portions in every system. As shown in a number of numerical studies (see, e.g., \citealt{brunet15,tiongco19}), dynamical evolution is expected to smooth out primordial kinematic and structural differences in the innermost regions first and then in the cluster’s outskirts. Therefore, capturing rotation differences between MPs require a
compromise between probing a fairly wide radial coverage, thus to trace regions where kinematic differences should be present 
for a longer time, and sampling distances from the cluster center where rotation is more prominent.  With this in mind, we decided to measure $\alpha$ within $R_{\rm \alpha}$. We verified also that the adoption of different radial selections does not have a significant impact on the overall relative distribution of $\alpha$ values.
Errors on $\alpha$ were obtained by propagating the posterior probability distributions obtained from the MCMC analysis for the best-fit rotation and velocity dispersion profiles derivation (see Section~3.1).
Differences between SP and FP kinematic patterns are constrained simply by ($\alpha^{\rm SP}-\alpha^{\rm FP}$). With such a  definition, a more rapidly rotating SP yields positive values of ($\alpha^{\rm SP}-\alpha^{\rm FP}$).

Following similar lines, we defined a parameter to describe the 3D rotation:
\begin{equation}\label{vel3}
\omega^X_{3D} = (A_{\rm 3D}/\sigma_0^{\rm 3D})/(R_{\rm m}/R_{\rm h})
\end{equation}
where $\sigma_0^{\rm 3D}$ represents the 3D central velocity dispersion and it is defined as the quadratic average of the $\sigma_0$ values obtained for the three velocity components (i.e., \texttt{LOS}, \texttt{TAN} and \texttt{RAD} - Section~3.1); $R_{\rm m}$ is the average cluster-centric distance of stars for which we have tri-dimensional velocity measures, and $R_{\rm h}$ is the system half-light radius (from \citealt{harris96}). 
Here $R_{\rm h}$ is adopted as a meaningful radial normalization factor to secure a direct comparison among different GCs attaining significantly different projected radial extension.
In the assumption of a pure solid-body rotation, $\omega_{\rm 3D}$ would represent the best-fit angular rotation.
As for the 1D analysis, the introduction of $\omega_{\rm 3D}$ is primarily meant to provide a direct and reliable characterization of the 3D rotation based only on quantities that are directly derived from the observations. Differences in the 3D rotation of SP and FP are given by ($\omega_{\rm 3D}^{\rm SP}-\omega_{\rm 3D}^{\rm FP}$), which yields positive values for a more rapidly rotating SP.

\begin{figure*}
\centering
\includegraphics[width=0.95\textwidth]{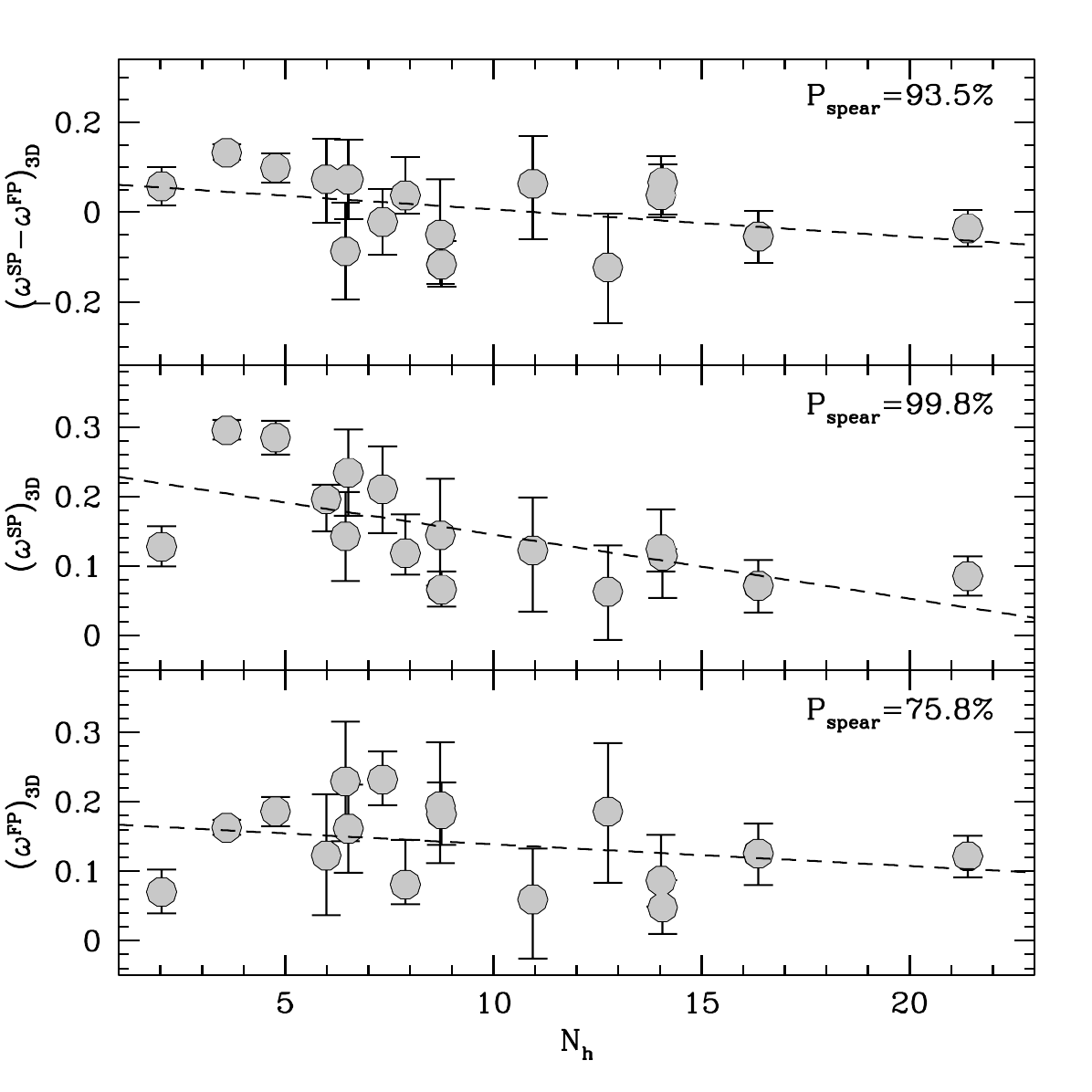}
\caption{As in Figures~\ref{fig:rot_mp} and \ref{fig:rot_mp_pm}, but now for the $\omega_{\rm 3D}$ parameter (Section~5.1).}
\label{fig:rot_mp_3D}
\end{figure*}

Several works have shown that the rotation strength observed in GCs is primarily shaped by their dynamical age, with dynamically young systems typically showing the larger degree of rotation (e.g. \citealt{fabricius14,bianchini18,kamann18,sollima19}).
We used $N_{\rm h}=t_{\rm age}/t_{\rm rh}$ as a proxy of the clusters' dynamical age. We adopted the 
the half-mass relaxation time ($t_{\rm rh}$) values reported by \citet{harris96} and
ages derived by \citet{dotter10} for all clusters but NGC~1904, for which we used the age inferred by \citet{dalessandro13}. 

In Figure~\ref{fig:rot_tot} in the Appendix we show the distribution of the $\alpha$ values (along both the \texttt{LOS} and \texttt{TAN} velocity components) and of $\omega_{\rm 3D}$ as a function of $N_{\rm h}$ 
for the TOT population. 
In general, we find a very good agreement with previous analysis (e.g., \citealt{bianchini18,kamann18,baumgardt19,sollima19}) in terms of correlation between cluster rotation strength and dynamical age, thus further strengthening the idea that the
present-day cluster rotation is the relic of that imprinted at the epoch of cluster formation, and that it has been then progressively dissipated via two-body relaxation \citep{einsel99,hong13,tiongco17,livernois22,kamlah22}.
Interestingly, such an agreement provides also an independent assessment about the reliability of the adopted kinematic parameters.

In Figures~\ref{fig:rot_mp}, \ref{fig:rot_mp_pm} and \ref{fig:rot_mp_3D} we show the distributions of 
$\alpha$ (for both the \texttt{LOS} and \texttt{PM} components) and of $\omega_{\rm 3D}$ as a function of $N_{\rm h}$ for both SP and FP stars (bottom and middle panels).
Interestingly, a number of common patterns can be highlighted in the three Figures. 
First, we note that in a large fraction of clusters (up to $\sim50\%$) both FP and SP show evidence of non-negligible rotation. Secondly, both sub-populations show evidence of anti-correlation with $N_{\rm h}$ in all three analyzed velocity components, with dynamically young clusters 
being characterized by a larger rotation strength. 
This behavior turns out to be more prominent when the PM and 3D analysis are considered. 
This is somehow expected as PMs are less affected than the \texttt{LOS} velocities by the smoothing introduced by the superposition of stars located at different cluster-centric distances and attaining different rotation velocities. In addition, the 3D 
analysis accounts for any rotation axis inclination and projection effects. 
Finally, we observe that for all velocity components, in dynamically young clusters the SP is characterized by larger $\alpha$ and $\omega_{\rm 3D}$ values 
(i.e., more rapid rotation) than that observed for the FP at similar $N_{\rm h}$, and it shows a more rapid decline 
than the FP as a function of $N_{\rm h}$. In fact, a Spearman correlation test gives a probability $P_{\rm spear}$ 
larger than $\sim99\%$ 
of correlations between $(\alpha^{\rm SP})_{\rm PM}$ or $\omega_{\rm 3D}^{\rm SP}$ and $N_{\rm h}$, 
while probabilities are smaller when either the \texttt{LOS} component or the FP is considered. 
We note here that by using the approach described in \citep{curran15} we have verified that
the results of the Spearman rank correlation tests performed in our analysis and reported in the following 
are robust against possible outliers and errors associated to the adopted kinematic parameters.

\begin{figure*}
\includegraphics[width=0.95\textwidth]{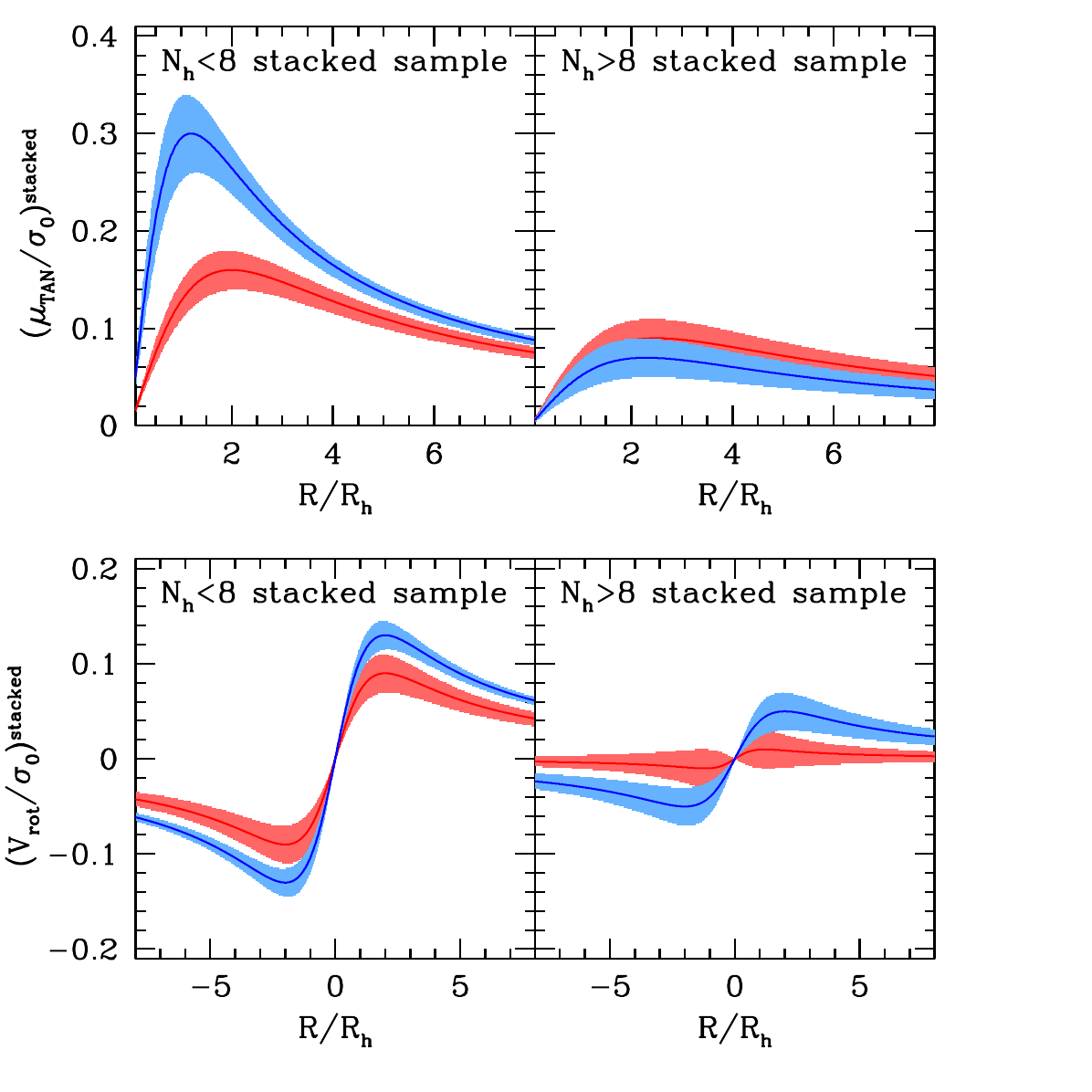}
\caption{Best-fit results of the kinematic analysis of the dynamically young ($N_{\rm h}<8$) and old ($N_{\rm h}>8$)
stacked samples. FP is in red and SP in blue. The lower panels refer to the \texttt{LOS} rotation, while the upper row shows the 
results for the \texttt{TAN} velocity component.}
\label{fig:rot_sc}
\end{figure*}

To better highlight such a differential behavior, the upper panels of Figures~\ref{fig:rot_mp}, \ref{fig:rot_mp_pm} and \ref{fig:rot_mp_3D} show the difference between the rotation strength of the SP and FP as given by ($\alpha^{\rm SP}-\alpha^{\rm FP}$) and 
($\omega_{\rm 3D}^{\rm SP}-\omega_{\rm 3D}^{\rm FP}$). 
Admittedly, nor ($\alpha^{\rm SP}-\alpha^{\rm FP}$) or
($\omega_{\rm 3D}^{\rm SP}-\omega_{\rm 3D}^{\rm FP}$) show striking variations 
in the dynamical age range sampled by the target clusters ($2<N_{\rm h}<25$). 
In fact, Spearman rank correlation tests give probabilities of correlation of $P_{\rm spear}\sim90\%$ ($\sim95\%$ in the 3D case). 
Interestingly however, a negative trend between the rotation strength differences and $N_{\rm h}$ is consistently observed in all velocity components. 
In fact, both ($\alpha^{\rm SP}-\alpha^{\rm FP}$) and
($\omega_{\rm 3D}^{\rm SP}-\omega_{\rm 3D}^{\rm FP}$)
show positive values for dynamically young GCs and then they progressively approach 0 for dynamical older clusters, meaning that FP and SP rotate at the same velocity. The nice agreement between the results obtained in the three analysis definitely provides support to the fact that there is a real correlation between the SP and FP rotation strength differences and $N_{\rm h}$, and that in general SP shows a more rapid rotation than the FP at dynamically young ages. 
These results represent the first observational evidence of the link between MP rotation patterns and clusters' long-term dynamical evolution.

We do not find any significant difference between the MP rotation axis orientation both for the \texttt{LOS} and 3D analysis.
In fact, the mean difference between the best-fit $PA_0$ values of the SP and FP is $-2^{\circ}\pm19^{\circ}$,
and those between $\theta_0$ and $i$ are $4^{\circ}\pm24^{\circ}$ and $2^{\circ}\pm12^{\circ}$, respectively.

%%%SUPER CLUSTER
As an additional way to analyze the data and search for possible trends, we divided the clusters in two sub-groups according to 
their dynamical ages. In particular, we defined a group of clusters with $N_{\rm h}<8$ and a complementary one ($N_{\rm h}>8$) 
including all the remaining GCs. In this way the two sub-groups turn out to be populated by the same number of systems. 
Within each sub-group and sub-population, we then stacked all the available kinematic information after normalizing the 
cluster-centric distances to the cluster's $R_{\rm h}$ (from \citealt{harris96}), the velocities to the central velocity 
dispersion in a given velocity component (as obtained by the analysis described in Section~3) and rotating all clusters to 
have the same $PA_0$ (Section~3.1). In this way, we could jointly compare the behavior of MPs
for multiple clusters at once, thus increasing the number of stars that can be used to study the kinematics of each sub-population and narrowing down the uncertainties on the 
derived kinematic parameters 
The kinematics of MPs in the two stacked samples was then analyzed by following the same approach described in Section~3 and previously 
adopted for single GCs.
Figure~\ref{fig:rot_sc} shows the results of the kinematic analysis for the two stacked samples for both the \texttt{LOS} and
\texttt{TAN} components (lower and upper panels, respectively). In both cases, a significant difference between the FP and SP rotation
profiles is observed for the dynamically young ($N_{\rm h}<8$) stacked sample, 
while they almost disappear for the dynamically 
old GCs. We then derived the same kinematic parameters described by Equation~9 for a direct comparison with single GCs.
Results are shown in Figures~\ref{fig:rot_mp} and \ref{fig:rot_mp_pm} by the two starry symbols. 
Both in the \texttt{LOS} and \texttt{TAN} components and for each sub-populations, 
results are fully consistent with the general trend described by single clusters. 
Interestingly, the reduced uncertainties strengthen the significance of the observed differences discussed above. 
In particular, the stacked analysis shows 
that the observed trends between $\alpha^{\rm FP}$, $\alpha^{\rm SP}$ and $N_{\rm h}$ along both the \texttt{LOS} and \texttt{TAN} 
components are significant at a large confidence level ($P_{\rm stacked}>5\sigma$). 
Also, while the ($\alpha^{\rm SP}-\alpha^{\rm FP}$) difference between dynamically young and old clusters is only marginally 
significant along the \texttt{LOS} component, it turns out to be significant at a $\sim6\sigma$ level for the tangential component.

Unfortunately, we could not apply the 3D rotation analysis (Section~3.3) to the stacked samples as it is not possible to report all clusters to the same values of $\theta_0$ and $i$.

\begin{figure}
\centering
\includegraphics[width=\columnwidth]{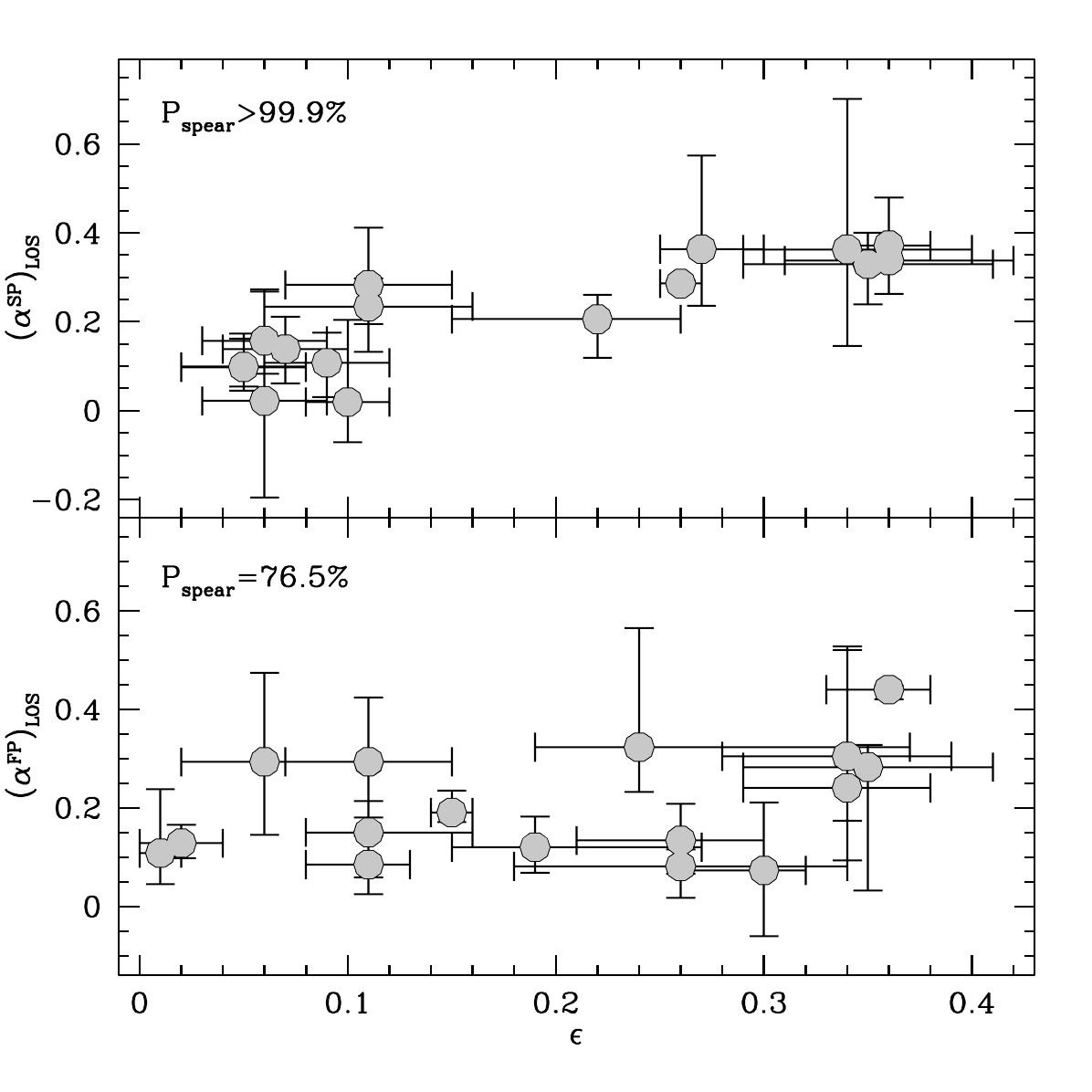}
\caption{Distributions of the $(\alpha)_{\rm LOS}$ parameter for the FP and SP (lower and upper panel respectively) as a function of the best-fit ellipticity values (Section~4) for the two sub-populations. }
\label{fig:ell_MP}
\end{figure}

\subsection{Ellipticity}
In general a rotating system is also expected to be flattened in the direction perpendicular to the rotation axis \citep{chandra69}. 
Under the assumption that GCs can be described by the same dynamical model, such as an isotropic oblate rotator (e.g., \citealt{varri12}), stronger rotation would be expected in more flattened systems. However, various effects can dilute a possible correlation, the most important ones being anisotropies, inclination effects or tidal forces from the Milky Way (see \citealt{vdb08}, for an estimate of the impact of the latter). 
Nevertheless, \citet{fabricius14} and \citet{kamann18} were able to reveal a correlation between cluster rotation and ellipticity in a sample of Galactic GCs (see also \citealt{lanzoni18a,dalessandro21a,leanza23} for similar analysis on specific clusters).

Following on those results, we searched for any link between MP rotation and ellipticity for all clusters in our sample.
The results of such a comparison for population TOT are reported in Section~\ref{Appendix B}.  
Interestingly, we find a clear and significant correlation between these two quantities (Spearman rank correlation test gives probabilities $\sim99.9\%$), in nice agreement with previous results by \citet{fabricius14} and \citet{kamann18}.

Here we explore in some detail whether also the MP rotation patterns are possibly linked to their morphological 2D spatial distributions. We compared the ellipticity estimates obtained as described in Section~4 with the $(\alpha)_{\rm LOS}$ MP values in Figure~\ref{fig:ell_MP}.
Both populations show a positive correlation between ($\alpha$)$_{\rm LOS}$ and $\epsilon$, with Spearman rank 
correlation probabilities $P_{\rm spear}\sim80\%$ and larger than $99.9\%$ for the FP and SP, respectively. 
In detail, the FP shows a pretty flat distribution of ($\alpha^{\rm FP}$)$_{\rm LOS}$ up to ellipticity 
values $\epsilon\sim0.2$. Then, ($\alpha^{\rm FP}$)$_{\rm LOS}$ starts to increase almost linearly 
with $\epsilon$. As discussed in Section~5.1, the SP tends to show larger values of rotation than the FP. 
Likely driven by such a stronger rotation, in the upper panel of Figure~\ref{fig:ell_MP}, we observe that ($\alpha^{\rm SP}$)$_{\rm LOS}$ follows a nicely linear correlation with $\epsilon$ for the entire range of ellipticity values sampled by the target GCs. 

Following the analysis by \citet{fabricius14} and \citet{kamann18} we computed the differences between the 
PA values obtained in Section~4 for the 2D stellar spatial distribution and the best-fit rotation axis 
position angles ($PA_0$). Interestingly, while the distribution of the difference is pretty scattered we find that the average value for the systems in our sample is $\sim85^{\circ}$, thus implying that the stellar density distribution is on average flattened in the direction perpendicular to the rotation axis. This behavior is in general agreement with what is expected for a rotating system and it is qualitatively consistent with what predicted, for example, by the models introduced by \citet{varri12} and previously found in other observational studies (e.g., \citealt{bianchini13,bellini17,lanzoni18a,dalessandro21a,leanza22}).

\begin{figure*}
\centering
\includegraphics[width=1\textwidth]{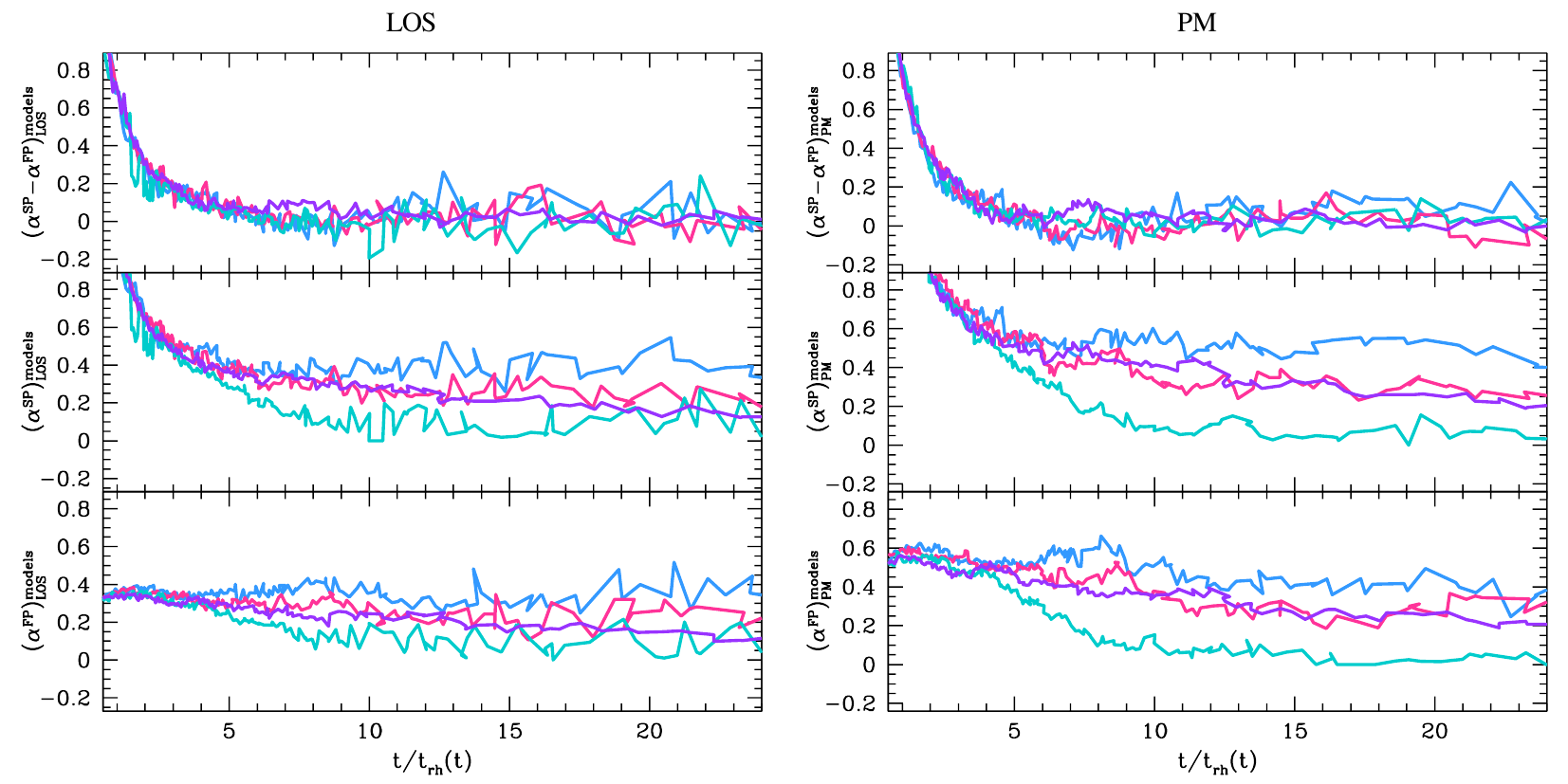}
\caption{Time evolution of the rotational parameters $\alpha^{\rm FP}$, $\alpha^{\rm SP}$, and their differences 
(see Section 5.1) for the simulations described in Section~5.3. The blue line corresponds to the model with $\delta=0^{\circ}$,
the pink, cyan and purple correspond to $\delta=45^{\circ},90^{\circ}$ and $180^{\circ}$, respectively.}
\label{fig:models_rot}
\end{figure*}

\subsection{Dynamical simulations}
To conclude the discussion about the rotational properties 
of MPs in our target GCs,
we briefly present the results of a set of $N$-body simulations aimed at exploring the evolution of rotating MP clusters. A full discussion and detailed description of the results of these simulations will be presented in White et al. (in prep.), here we just report the evolutionary path followed by the $\alpha$ parameter 
introduced in this paper to trace the strength of rotation of FP and SP stars and their difference.
In our simulations we focus only on the long-term dynamics driven by the effects of two-body relaxation for star clusters evolving in the external tidal field of their host galaxy. Each system starts with $10^5$ stars with masses between 0.1 and 1 $M_\odot$ distributed according to a \citet{kroupa01} stellar initial mass function. Our systems start with an equal number of FP and SP stars; following the general properties emerging from a few studies of the formation of SP stars in rotating clusters (see e.g., \citealt{bekki10,bekki11,lacchin22}) the SP is initially more centrally concentrated and more rapidly rotating than the FP. The two populations rotate around a common axis. To explore the interplay between internal dynamics and the effects due to the external tidal field we have explored models with different angles $\delta$ between the internal rotation axis and the rotation axis of the cluster orbital motion around the center of the host galaxy. In particular, we have explored systems with values of $\delta$ equal to $0^{\circ}$, $45^{\circ}$, $90^{\circ}$, and $180^{\circ}$. The simulations were run with the {\tt NBODY6++GPU} code \citep{wang15}.
In Fig.~\ref{fig:models_rot} we show the time\footnote{Time is normalized to the half-mass relaxation time 
$t_{rh}= 0.138 M^{1/2}r_h^{3/2}/(G^{1/2}\ln(0.02N)\langle m \rangle)$ where $r_{\rm h}$ is the cluster half-mass radius, 
$M$ is its total mass, $N$ the total number of stars, and $<m>$ is the mean stellar mass} evolution of $\alpha^{\rm FP}$, $\alpha^{\rm SP}$ and ($\alpha^{\rm SP}-\alpha^{\rm FP}$) for these models using rotational velocity profiles calculated for both the \texttt{LOS} (left panel) and the \texttt{TAN} component (right panel).
For these plots we adopt an ideal line of sight perpendicular to the cluster angular momentum or parallel to it. We emphasize that these simulations are not aimed at a detailed comparison with observations but they rather serve as a guide to illustrate the extent of the effects of dynamical processes on the initial differences between the rotational kinematics of the FP and SP populations. We also reiterate that our simulations are focussed on the effects of two-body relaxation and do not include early dynamical phases such as those during which a star cluster responds to the mass loss due to stellar evolution, which can have an effect on the sub-populations' dynamical differences (see e.g. \citealt{vesperini21,sollima21}).

By construction, the $\alpha$ values derived for the SP are significantly larger by about a factor of 2-3 than those of the FP. The results of our simulations show that the effects of two-body relaxation lead to a rapid and significant reduction of the initial difference between the FP and the SP rotation in the first 2-3 half-mass relaxation times reaching values of ($\alpha^{\rm SP}-\alpha^{\rm FP}$) similar to those found in our observational analysis. Then at later dynamical ages, ($\alpha^{\rm SP}-\alpha^{\rm FP}$) keeps decreasing at a lower pace and it progressively approaches values close to 0 around 10 relaxation times, when FP and SP stars rotate at the same velocity. 
We note that, as already discussed in Section 5.1 and in agreement with what was found in the observations, the rotation strength for both the FP and SP along with their difference is stronger when a PM-like projection is considered (Fig.~\ref{fig:models_rot}).
The behavior described by the simulations is in nice general agreement with the observed trends (Figures~\ref{fig:rot_mp}, \ref{fig:rot_mp_pm} and \ref{fig:rot_mp_3D}). Such an agreement strongly suggests that both the rotational differences and the mild trend between the rotation strength and the dynamical age revealed by our observational analysis are consistent with those expected for the long-term dynamical evolution of GCs born with a SP initially rotating more rapidly than the FP. 

\begin{figure*}
\centering
\includegraphics[width=0.95\textwidth]{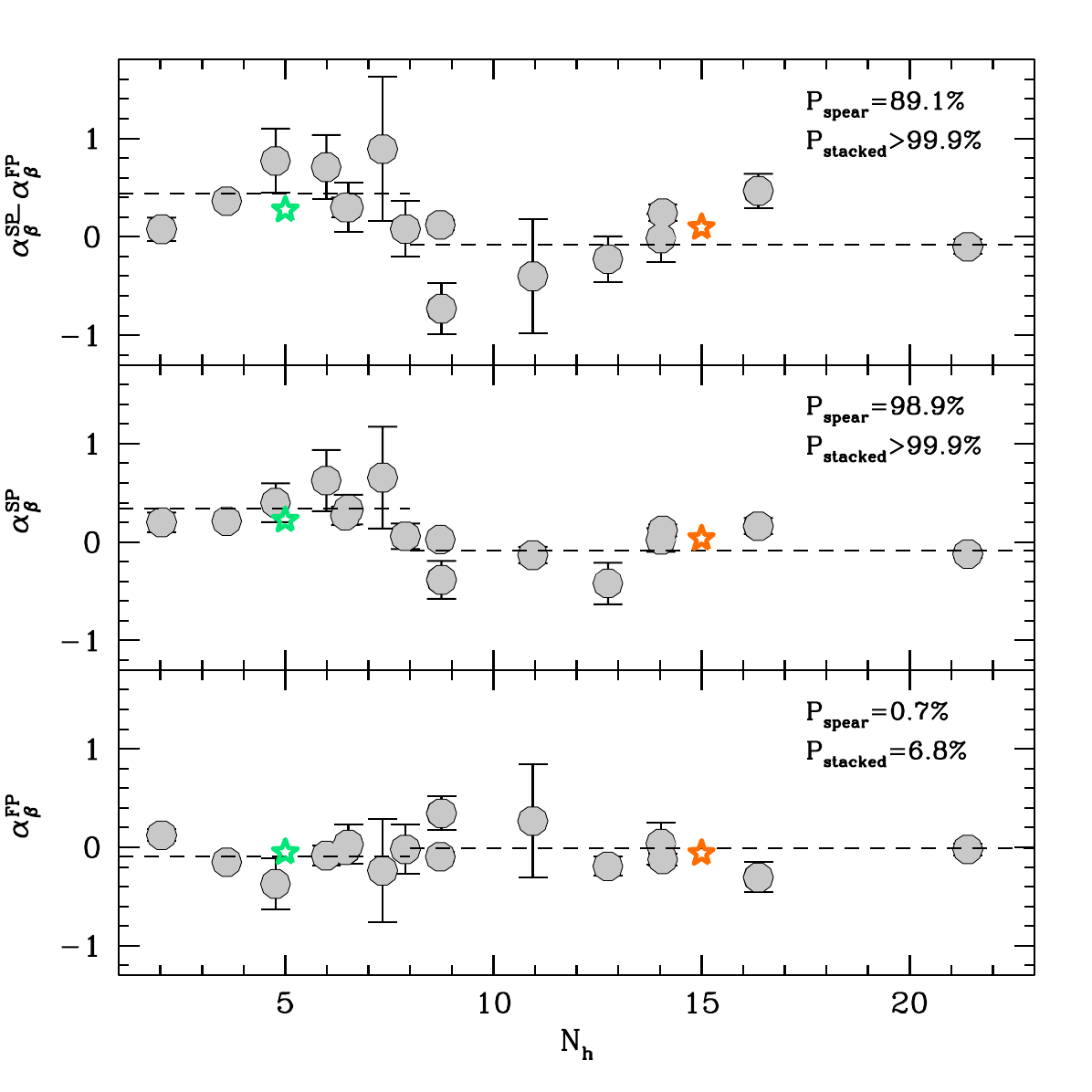}
\caption{The bottom and middle panels show the distribution of the anisotropy parameter $\alpha_{\rm \beta}$ for the FP and SP sub-populations of GCs in the sample (grey circles) as a function of $N_{\rm h}$. The top panel, shows the differential trend $\alpha_{\rm \beta}^{\rm SP} - \alpha_{\rm \beta}^{\rm FP}$. The dashed lines represent the mean of the observed $\alpha_{\rm \beta}$ values derived for the dynamically young ($N_{\rm h}<8$) and old ($N_{\rm h}>8$) GCs. 
The starry symbols are the result of the analysis on the two stacked samples and their sizes correspond to the errorbars. 
}
\label{fig:ani}
\end{figure*}

\section{Results: anisotropy}
\subsection{Observations}
We investigated the differences in the mean anisotropy properties of MPs by defining the following parameter:

\begin{equation}
\alpha_{\rm \beta}=\int_0^{1} (\beta(R_l)_{X}) dR_l
\end{equation}

here $R_l$ represents the distance from the cluster center normalized to the value of the clusters' Jacobi radius (from \citealt{baumgardt19}) to allow a meaningful comparison among different systems. From a geometrical point of view, Equation~11 represents the area subtended by the best-fit radially normalized anisotropy profile (as defined in Section~3.2). By definition, if a sub-population is radially anisotropic, $\alpha_{\rm \beta}$ will have positive values, while it will attain negative values in case of tangential anisotropy. Differences between the anisotropy properties of MPs are given by ($\alpha_{\rm \beta}^{SP} - \alpha_{\rm \beta}^{FP}$). As before, errors on $\alpha_{\rm \beta}$ 
were obtained by propagating the posterior probability distributions obtained from the MCMC analysis for the best-fit anisotropy profiles.

\begin{figure*}
\includegraphics[width=0.95\textwidth]{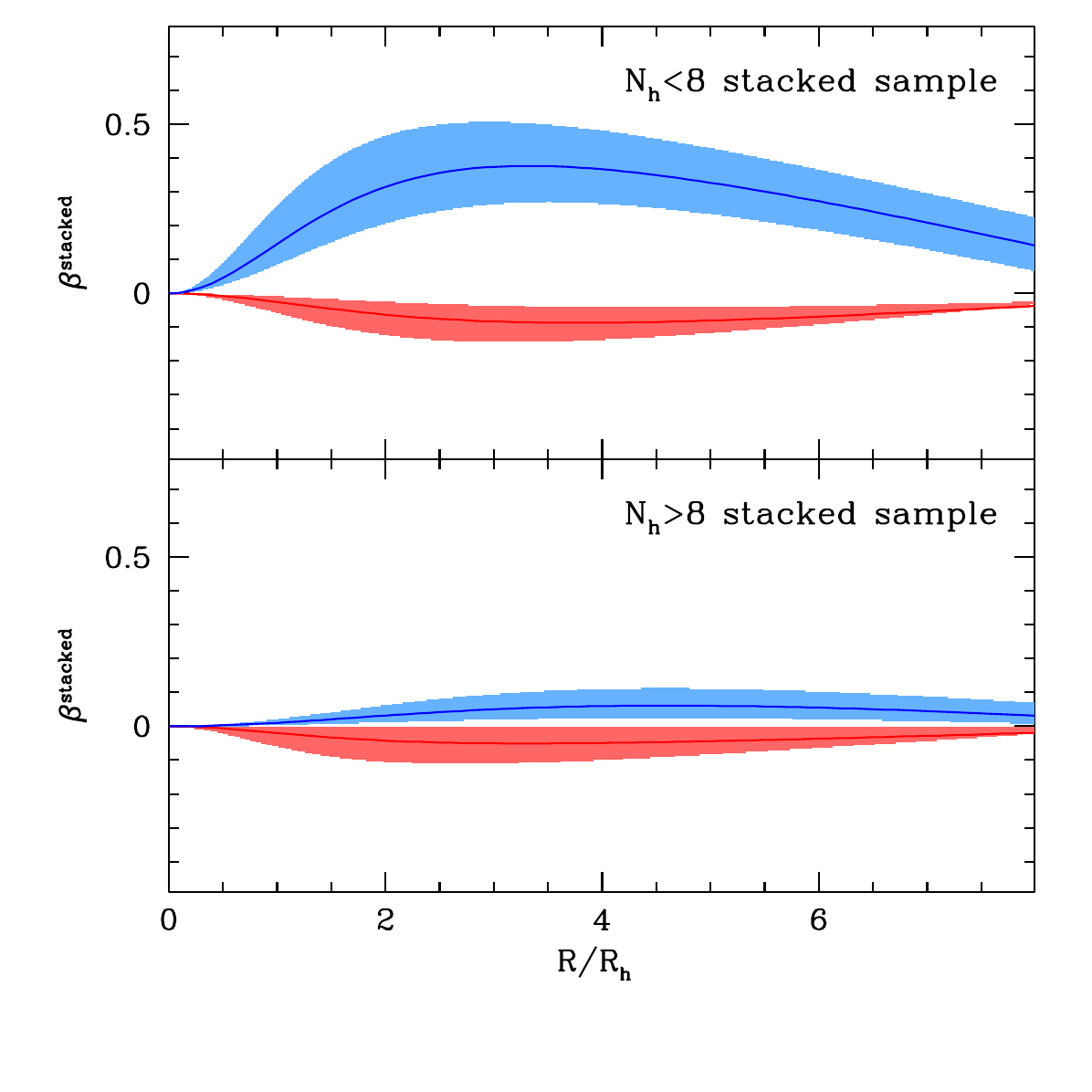}
\caption{Best-fit results of the anisotropy analysis of MPs in the two stacked samples. The upper panels shows the best-fit 
anisotropy profiles of the FP (red) and SP (blue) sub-populations as obtained for the stacked sample of dynamically young GCs, while
the lower panel refers to dynamically old systems.}
\label{fig:ani_stacked}
\end{figure*}

Similarly to the rotation analysis, Figure~\ref{fig:ani} shows the variation of $\alpha_{\rm \beta}$ as a function of the dynamical age for the FP and SP in the bottom and middle panel, while the distribution of ($\alpha_{\rm \beta}^{SP} - \alpha_{\rm \beta}^{FP}$) is shown in the top panel. 
$\alpha_{\rm \beta}^{FP}$ shows a flat distribution around $\sim0$ at any $N_{\rm h}$, thus suggesting that the FP is on average isotropic for all GCs in the sample, independently on their dynamical ages. On the contrary, the $\alpha_{\rm \beta}^{SP}$ shows a non negligible variation as a function of time. 
In fact, for dynamically young clusters, $\alpha_{\rm \beta}^{SP}$ is positive (i.e., SP is radially anisotropic) and 
it becomes isotropic only for the dynamically older systems. To better highlight such a trend, we can use the two GC 
sub-groups defined in Section~5.1. At a face value, GCs with $N_{\rm h}<8$ (dynamically young) have a mean 
$\alpha_{\rm \beta}^{SP}$ value of $0.40\pm0.20$, while for dynamically older clusters the mean value is $-0.09\pm0.23$ (dashed lines in Figure~\ref{fig:ani}). 
Such a difference is very nicely described by the comparison between the best-fit results obtained 
for MPs in the two stacked samples as shown in Figure~\ref{fig:ani_stacked}.
Indeed, when the anisotropy analysis is performed on the two stacked sub-samples, 
we find that $\alpha_{\rm \beta}^{SP}=0.23\pm0.03$ at dynamically young ages, while for dynamically older 
clusters we obtain $\alpha_{\rm \beta}^{SP}=0.04\pm0.03$, thus implying that the difference in the average 
SP orbital anisotropy is significant at a $\sim 5\sigma$ level.
For comparison, we computed the same quantities for the FP on the stacked clusters. As expected, in this case 
mean values are virtually indistinguishable within the errors.
The different behavior between the average orbital properties of FP and SP stars is further highlighted by the 
distribution of ($\alpha_{\rm \beta}^{SP} - \alpha_{\rm \beta}^{FP}$). In dynamically young GCs SP is systematically 
more radial anisotropic than the FP (mean difference being $0.44\pm0.31$), then they both become compatible with 
being isotropic at later dynamical stages ($-0.08\pm0.37$). 
The analysis on the two stacked sub-samples provides values a 
($\alpha_{\rm \beta}^{SP} - \alpha_{\rm \beta}^{FP}$)$=0.28\pm0.07$ and $0.09\pm0.07$ for the dynamically young 
and old groups, respectively, thus implying that there is a real orbital velocity distribution difference between 
FP and SP stars with a significance of $\sim3.5\sigma$.

\subsection{Dynamical models}
\begin{figure}
\centering
\includegraphics[width=\columnwidth]{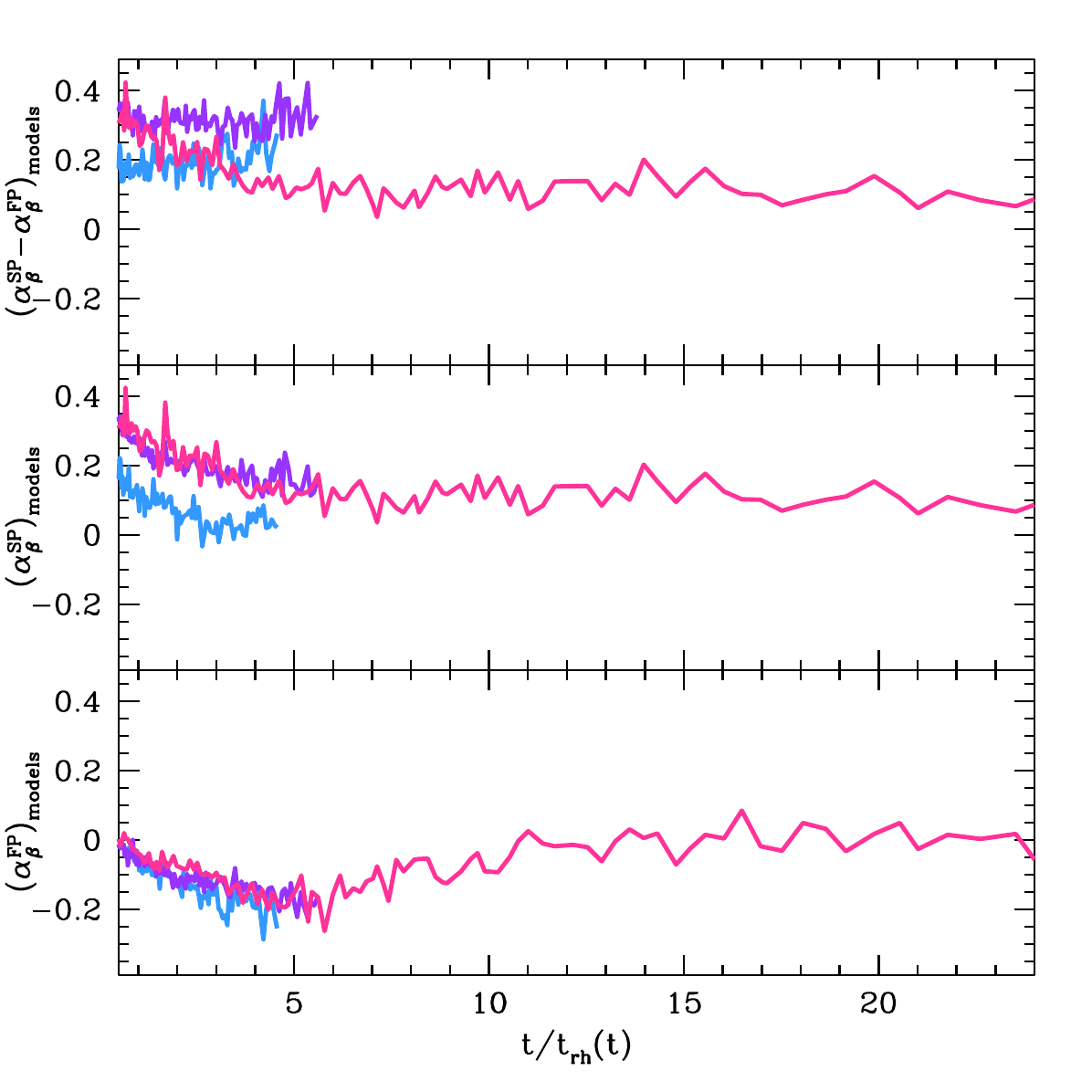}
\caption{Time evolution of the anisotropy parameters defined in Section~6.1 for the FP (bottom panel), the SP (middle panel) and their difference
(top panel) in the three Monte Carlo simulations described in Section 6.2 (N1Mr10 blue line, N1M purple line; N05M pink line).
Time is normalized by the half-mass relaxation time.}
\label{fig:models_ani}
\end{figure}

The difference between the anisotropy of the FP and SP populations is consistent with the predictions of a number of theoretical investigations of the dynamics of MP clusters. These studies find that in clusters where the SP forms more centrally concentrated than the FP, the SP is generally characterized by a more radially anisotropic velocity distribution than the FP (see e.g., \citealt{bellini15,brunet15,tiongco19,vesperini21,sollima21}). In order to illustrate the 
expected strength and evolution of the differences between the anisotropy of the FP and SP populations, as measured by the parameter ($\alpha_{\rm \beta}^{SP}-\alpha_{\rm \beta}^{FP}$) introduced in this paper, we show in Figure~\ref{fig:models_ani} the time evolution of  $\alpha_{\rm \beta}^{FP}$, $\alpha_{\rm \beta}^{SP}$ and ($\alpha_{\rm \beta}^{SP}-\alpha_{\rm \beta}^{FP}$) for three Monte Carlo simulations following the dynamical evolution of MP clusters with different initial conditions. The Monte Carlo simulations were run with the {\tt MOCCA} code 
(\citealt{hypki13,giersz13}; see also \citealt{vesperini21,hypki22} for the specific case of GCs with MPs).
A full presentation of the kinematics and phase space properties of these systems will be presented in a separate paper (Aros et al. in prep.). The three models for which we present the anisotropy evolution in Fig.~\ref{fig:models_ani} all start with a ratio of 
the SP to total mass equal to $0.2$. The total number of stars is equal to $N=10^6$ (hereafter we use the id. {\it N1M} for these model) or $0.5\times 10^6$ (with id. {\it N05M}) and stars have been extracted from a \citet{kroupa01} initial stellar mass function between 0.1 and 100 $M_{\odot}$. The SP is initially more concentrated than the FP: for two models the initial ratio for the FP to SP half-mass radii is equal to 20 ({\it N1Mr20} and {\it N05Mr20}) and is equal to 10 for one model ({\it N1Mr10}). Both populations are initially characterized by a non-rotating, isotropic velocity distribution. All models include the effects of mass loss due to stellar evolution, two-body relaxation, binary interactions and a tidal truncation (see \citealt{vesperini21} for further details on the evolutionary processes included in these simulations).
As shown in Figure~\ref{fig:models_ani}, in all models the SP is characterized by a more radially anisotropic velocity distribution than the FP. Even if both populations are initially isotropic, the development of a stronger radial anisotropy for the SP is the consequence of its initial more centrally concentrated spatial distribution (see e.g., \citealt{tiongco19,vesperini21}). 
The differences in anisotropy between the two populations gradually decrease as the systems evolves. As shown by the results of our simulations, the extent of the differences between the FP and SP anisotropy depends on the initial relative concentration of the two populations being smaller for the model where the initial ratio of the FP to SP half-mass radii equal to 10 ({\it N1MR10}) than that for models where this ratio is initially equal to 20 ({\it N1MR20}).
While also in this case we point out that these simulations have not been specifically designed to provide a detailed fit to observations but rather to provide guidance in the general interpretation of the observational results, it is interesting to note that the extent of the difference between the anisotropy of the FP and the SP populations as measured by the ($\alpha_{\rm \beta}^{SP}-\alpha_{\rm \beta}^{FP}$) and its time evolution is generally consistent with the findings of our observational analysis (Figure~\ref{fig:ani}). The general agreement between observations and simulations lends further support to the interpretation of the observed difference between the FP and SP anisotropy and its variation with the clusters' dynamical age as the manifestation of the different dynamical paths followed by subsystems with different initial spatial distributions.

\begin{table*}
\centering
\caption{MP kinematic parameters describing the radial anisotropy and rotation along the LOS, PM and 3D components.}
\label{PHOT_PAR}
\begin{tabular}{lccccc}
\hline \hline 
Cluster & $(\alpha)_{LOS}$ & $(\alpha)_{PM}$ & $\omega_{3D}$ & $\alpha_{\rm \beta}$ & POP \\
\hline
NGC 104 (47 Tuc) &  $0.19^{+0.04}_{-0.02}$ & $0.34^{+0.08}_{-0.06}$ & $0.16^{+0.02}_{-0.02}$  & $-0.15\pm0.04$ & FP   \\
                 &  $0.28^{+0.01}_{-0.01}$ & $0.51^{+0.06}_{-0.05}$ & $0.30^{+0.02}_{-0.02}$  & $0.21\pm0.01$ & SP \\
                 &    &    &    &   & \\  
NGC 288          &  $0.12^{+0.06}_{-0.05}$  &  $0.52^{+0.30}_{-0.27}$ & $0.12^{+0.09}_{-0.09}$ & $-0.08\pm0.27$ & FP \\
                 &  $0.14^{+0.07}_{-0.08}$  &  $0.72^{+0.02}_{-0.36}$ & $0.20^{+0.02}_{-0.05}$ & $0.63\pm0.31$ & SP \\ 
                 &    &    &    &   & \\ 
NGC~1261         &  $0.32^{+0.24}_{-0.09}$  & $0.22^{+0.45}_{-0.36}$ & $0.19^{+0.09}_{-0.08}$  & $-0.10\pm0.05$ & FP  \\
                 &  $0.36^{+0.21}_{-0.13}$  & $0.32^{+0.18}_{-0.26}$ & $0.14^{+0.08}_{-0.07}$ & $0.03 \pm 0.01$ & SP\\
                 &    &    &    &   & \\ 
NGC 1904 (M 79)  &  $0.29^{+0.18}_{-0.15}$  & $0.21^{+0.22}_{-0.22}$ & $0.08^{+0.06}_{-0.4}$  & $0.04 \pm 0.21$ & FP   \\
                 &  $0.16^{+0.12}_{-0.07}$  & $0.57^{+0.18}_{-0.47}$ & $0.12^{+0.06}_{-0.03}$ & $0.02 \pm 0.16$ & SP \\
                 &    &    &    &   & \\ 
NGC 3201         &  $0.07^{+0.14}_{-0.13}$  & $0.15^{+0.20}_{-0.08}$ & $0.23^{+0.08}_{-0.08}$ & $-0.03 \pm 0.10$ & FP     \\
                 &  $0.02^{+0.18}_{-0.09}$  & $0.14^{+0.18}_{-0.02}$ & $0.14^{+0.06}_{-0.06}$ & $0.27 \pm 0.09$ & SP \\ 
                 &    &    &    &   & \\ 
NGC 5272 (M 3)   &  $0.08^{+0.03}_{-0.06}$  & $0.34^{+0.15}_{-0.25}$ & $0.07^{+0.03}_{-0.03}$ & $0.12 \pm 0.06$ & FP   \\
                 &  $0.21^{+0.05}_{-0.09}$  & $0.35^{+0.14}_{-0.10}$ & $0.13^{+0.03}_{-0.03}$ & $0.20 \pm 0.10$ & SP \\
                 &    &    &    &   & \\ 
NGC 5904 (M 5)   &  $0.44^{+0.01}_{-0.02}$  & $0.44^{+0.22}_{-0.16}$ & $0.19^{+0.02}_{-0.02}$  & $-0.37 \pm 0.25$ & FP    \\
                 &  $0.37^{+0.01}_{-0.02}$  & $0.64^{+0.10}_{-0.15}$ & $0.29^{+0.02}_{-0.02}$ & $0.40 \pm 0.20$ & SP \\ 
                 &    &    &    &   & \\ 
NGC 5927         &  $0.11^{+0.13}_{-0.06}$  & $0.05^{+0.18}_{-0.06}$ & $0.05^{+0.04}_{-0.04}$ & $-0.12 \pm 0.06$ & FP \\  
                 &  $0.02^{+0.24}_{-0.22}$  & $0.07^{+0.25}_{-0.07}$ & $0.11^{+0.01}_{-0.06}$ & $0.12 \pm 0.06$ & SP \\ 
                 &    &    &    &   & \\ 
NGC 5986         &  $0.28^{+0.04}_{-0.25}$  & $0.30^{+0.34}_{-0.28}$ & $0.18^{+0.05}_{-0.04}$  &  $0.34 \pm 0.17$ & FP   \\
                 &  $0.33^{+0.07}_{-0.09}$  & $0.25^{+0.21}_{-0.20}$ & $0.07^{+0.3}_{-0.03}$  & $-0.38 \pm 0.19$ & SP \\
                 &    &    &    &   & \\ 
NGC 6093 (M 80)  &  $0.29^{+0.13}_{-0.15}$  & $0.29^{+0.13}_{-0.15}$ & $0.12^{+0.03}_{-0.03}$  & $-0.02 \pm 0.11$ & FP  \\
                 &  $0.28^{+0.13}_{-0.15}$  & $0.27^{+0.12}_{-0.14}$ & $0.09^{+0.03}_{-0.03}$ & $-0.12 \pm 0.20$ & SP \\
                 &    &    &    &   & \\ 
NGC 6171 (M 107) &  $0.37^{+0.08}_{-0.09}$ & $0.13^{+0.30}_{-0.13}$  & $0.19^{+0.10}_{-0.10}$ & $-0.19 \pm 0.09$ & FP \\
                 &  $0.44^{+0.10}_{-0.06}$ & $0.16^{+0.20}_{-0.03}$  & $0.06^{+0.07}_{-0.07}$ & $-0.41 \pm 0.21$ & SP \\
                 &    &    &    &   & \\ 
NGC 6205 (M 13)  &  $0.24^{+0.28}_{-0.15}$ & $0.24^{+0.28}_{-0.15}$  & $0.16^{+0.06}_{-0.06}$ & $0.03 \pm 0.20$ & FP \\    
                 &  $0.34^{+0.14}_{-0.07}$ & $0.34^{+0.14}_{-0.07}$  & $0.24^{+0.06}_{-0.06}$ & $0.33 \pm 0.15$ & SP \\
                 &    &    &    &   & \\ 
NGC 6362         &  $0.13^{+0.04}_{-0.03}$  & $0.13^{+0.04}_{-0.03}$  & $0.08^{+0.06}_{-0.03}$ & $-0.02 \pm 0.29$ & FP \\
                 &  $0.24^{+0.06}_{-0.04}$  & $0.24^{+0.06}_{-0.04}$  & $0.12^{+0.06}_{-0.03}$ & $0.06 \pm 0.13$ & SP \\
                 &    &    &    &   & \\ 
NGC 6254 (M 10)  &  $0.15^{+0.06}_{-0.09}$ &  $0.07^{+0.16}_{-0.20}$  & $0.13^{+0.04}_{-0.05}$ & $-0.31 \pm 0.15$ & FP \\
                 &  $0.11^{+0.07}_{-0.08}$ &  $0.03^{+0.15}_{-0.36}$  & $0.07^{+0.04}_{-0.04}$ & $0.16 \pm 0.08$ & SP \\
                 &    &    &    &   & \\ 
NGC 6496         &  $0.30^{+0.22}_{-0.13}$ &  $0.20^{+0.24}_{-0.15}$  & $0.06^{+0.07}_{-0.09}$ & $0.27 \pm 0.58$ & FP \\
                 &  $0.36^{+0.34}_{-0.22}$ & $0.28^{+0.16}_{-0.25}$  & $0.12^{+0.08}_{-0.09}$ & $-0.13 \pm 0.18$ & SP \\
                 &    &    &    &   & \\ 
NGC 6723         &  $0.13^{+0.07}_{-0.07}$ & $0.03^{+0.32}_{-0.29}$  & $0.23^{+0.04}_{-0.04}$ & $-0.24 \pm 0.52$ & FP\\
                 &  $0.10^{+0.08}_{-0.07}$ & $0.52^{+0.17}_{-0.04}$  & $0.21^{+0.06}_{-0.06}$ & $0.66 \pm 0.52$ & SP \\
\hline                 
\end{tabular}
\end{table*}

\begin{figure*}
\centering
\includegraphics[width=0.95\textwidth]{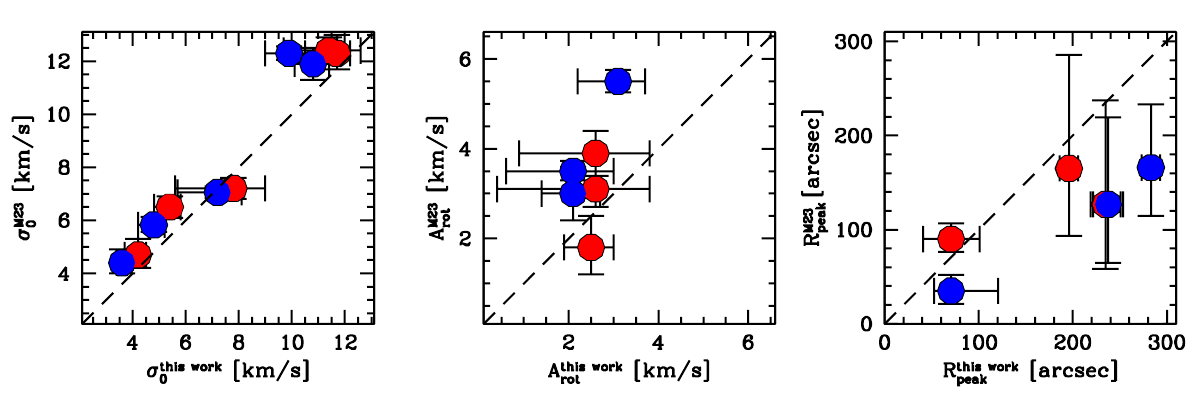}
\caption{Comparison between the best-fit $\sigma_0$, $A_{\rm rot}$ and $R_{\rm peak}$ values obtained for MPs in the present-work and by \citet{martens23}. Blue dots refer to SP and red ones to FP results.}
\label{fig:martens}
\end{figure*}

\section{Comparison with the literature}
In Appendix~\ref{Appendix C} we report on a quantitative comparison between the results obtained in the present work and those available in literature for the TOT population in each cluster. Here we detail on the comparison with previous works focusing on the kinematics of MPs.
We stress that the full 3D kinematic analysis presented in this paper is the first ever obtained for MPs. Hence, in the following our comparison will be limited to studies considering a single velocity component.  

Our sample has 6 GCs in common with the recent analysis by \citet{martens23} based on MUSE \texttt{LOS} velocities. 
The authors were able to find MP best-fit solutions for both the velocity dispersion and rotation profiles for three of them (namely 47~Tuc, NGC~5904 and NGC~6093), while for NGC~3201 and NGC~6254 they report only conservative upper limits for the MP rotation amplitudes, and for NGC~1904 they provide information only for the TOT population.
Figure~\ref{fig:martens} shows the distributions of the differences in terms of $\sigma_0$ (which is defined as $\sigma_{\rm max}$ in \citealt{martens23}), $A_{\rm rot}$ ($v_{\rm max}$ in \citealt{martens23}) and $R_{\rm peak}$ for the FP and SP sub-populations.
A good agreement is observed both for the MP central velocity dispersion values 
(left panel of Figure~\ref{fig:martens}) and the rotation amplitude (middle panel), with the most discrepant value being that corresponding to the rotation of the SP in 47~Tuc. 
Within the errors there is also a reasonable match between the values of $R_{\rm peak}$. We note however that, while the sample is certainly small, 
the estimates of $R_{\rm peak}$ by \citet{martens23} tend to be slightly smaller than those derived in this work. 
This might be somehow linked to the smaller radial coverage of the \citet{martens23} analysis. 
In fact, all values derived by \citet{martens23} for the clusters in common are located well outside the MUSE field of view and therefore they might be only partially constrained by their analysis. 

The sample analyzed in this work counts also 4 GCs in common with \citet{cordoni20}, namely 47~Tuc, NGC~288, NGC~5904 and NGC~6254. Our results are in agreement with theirs in that 47~Tuc and NGC~5904 are the systems showing the larger rotation among the clusters in common. However, at odds with their results, we also find that in both GCs the SP shows a larger rotation (as inferred both by $A_{\rm rot}$ and $\alpha$ values) than the FP. Also, within the uncertainties, we do not find evidence of any significant mis-alignment of the FP and SP rotation curves for these systems, neither in terms of position angles and inclination, as constrained by both the \texttt{LOS} and the full 3D analysis.

Finally, our analysis is in nice agreement with the results presented by \citet{cordero17} for the MP rotation patterns of M~13. 

As for the anisotropy, the results obtained in this work are in qualitative agreement with those found for the GCs M~13, NGC~2808 and 47~Tuc by \citet{richer13,bellini15,milone18}.
More in general, they are in nice qualitative agreement with those recently obtained in the innermost regions (R$<100\arcsec$) for a sample of Galactic GCs by \citet{libralato23} by using HST PMs to search for average differences between the anisotropy profiles of MPs.

\section{Summary and conclusions}
We have presented the first self-consistent 3D kinematic analysis of MPs for a sample of Galactic GCs. 
The study targets 16 systems spanning a broad range of dynamical ages ($2<N_{\rm h}<25$) and it is based 
on a large and mostly homogeneous observational dataset securing several hundreds of accurate \texttt{LOS} velocities and PMs for each 
cluster and sampling virtually their entire extension.

Our study is mainly focused on the analysis of the MP rotation along the three velocity components and the anisotropy of their velocity 
distributions. 
The adopted approach is aimed at providing new insights into the long term evolution of the kinematic properties of MPs 
(and their differences) for the entire sample of GCs and for the entire dynamical age covered by our analysis instead 
of focusing on the kinematic differences in specific clusters.
To this aim, starting from the observed velocity distributions we defined a few key quantities, to quantitatively and homogeneously compare the results obtained for all the observed GCs.

Our analysis provides the first observational determination of the dynamical path followed by MP kinematic properties during their long-term evolution.
The main observational results we find can be schematically summarized as follows.
\begin{itemize}
    \item {We observe evidence of differential rotation between MPs with the SP preferentially rotating more rapidly than the FP. This result is consistent (although with different amplitudes) along both the \texttt{LOS} and \texttt{TAN} velocity components, as well as in the full 3D analysis. In all GCs in our sample we find that the rotation position and inclination angles are consistent within the uncertainties between FP and SP.} 
    \item{The strength of the rotation signal of both FP and SP sub-populations nicely correlate with the ellipticity values derived for the two sub-populations. In addition, we find that the rotation axis position angle is typically perpendicular to the ellipses major axis.}
    \item{The difference in the rotation strength between MPs 
    is mildly anti-correlated with the cluster dynamical age. In particular, differences are larger for dynamically young clusters and they become progressively indistinguishable as dynamical evolution proceeds.}
    \item{Stars belonging to different sub-populations show different average velocity distribution properties. FPs are characterized by isotropic velocity distributions at any dynamical age probed by our sample. On the contrary, the SP of dynamically young GCs have a radially anisotropic velocity distribution which then becomes isotropic in more advanced evolutionary stages.}
\end{itemize}

The combination of these results with the analysis of the MP radial distributions of a representative sample of GCs carried out in our previous paper \citep{dalessandro19}, provides a full picture of the present-day kinematic and structural properties of MPs in GCs and of their evolution.
The comparison with dynamical models following the long-term evolution of MPs in GCs, suggests that these properties, 
and their evolution with dynamical age are in general good agreement with those expected in clusters forming with a SP subsystem initially more centrally concentrated and more rapidly rotating than the FP 
(see e.g., \citealt{dercole08,bekki10,calura19,lacchin22}). In turn, this would possibly suggest (see e.g., \citealt{brunet15} and discussion in \citealt{martens23}) that GCs experienced multiple events of star formation during their early phases of evolution, with the rotation properties being the more stringent discriminating factors. 
In fact, according to multi-epoch formation models, the SP is expected to form a low-mass, more centrally concentrated and more rapidly rotating stellar sub-system than the FP. In such a configuration, 
dissipative accretion processes of material ejected by FP stars and angular momentum conservation in sub-systems with different initial concentration can produce a larger initial rotation of the SP sub-system \citep{bekki11,brunet15,tiongco17}. Interestingly, it has been shown (e.g., \citealt{bekki11}) that even if only a very small fraction of the kinetic energy of the FP is in the form of bulk rotation energy, SP stars can acquire a much stronger rotation than what remains in the FP. 

It is important to note that, as shown in this paper and in a number of previous studies, 
early and long-term evolution can significantly reduce the initial 
differences between the FP and SP rotational velocities making them virtually indistinguishable in dynamical old systems.
The combination of these physical effects with the observational uncertainties arising from the limited available stellar samples, partial cluster coverage and possible biases introduced by the different rotation inclination angles, can make extremely difficult to capture present-day kinematic differences between MPs, in particular in the \texttt{LOS} velocity component. 
As a consequence, it is important to use some caution in drawing conclusions about the physical mechanisms at the basis of GC formation and early evolution based 
on the present-day kinematic and structural properties of individual systems or small samples. 

An homogeneous combination between data obtained with multi-object spectrographs and {\it {\it Gaia}} (as presented in this paper) which are particularly sensitive to the intermediate and large cluster-centric distances in GCs, with Integral Field Unit \texttt{LOS} RVs and HST absolute PMs sampling more efficiently the clusters' innermost regions (see for example \citealt{martens23,libralato23}), along with an increase of the cluster sample size, represents a natural and promising next step that could potentially enable the exploration of additional physical ingredients at play.

The results of this study clearly demonstrates that significant advances in our understanding of cluster formation and early evolution is only possible through a multi-faceted and multi-diagnostic approach and by combining state-of-art observations and simulations. 

\begin{acknowledgements}
The authors thank the anonymous referee for the careful reading of the paper and the useful comments that improved the presentation of this work.
E.D. acknowledges financial support from the Fulbright Visiting Scholar program 2023. 
E.D. and A.D.C. are also grateful for the warm hospitality of the Indiana University where part of this work was performed.
E.D. and M.C. acknowledge financial support from the project Light-on-Dark granted by MIUR through PRIN2017-2017K7REXT. 
E.V. acknowledges support from NSF grant AST-2009193. E.V. acknowledges also support from the John and A-Lan Reynolds Faculty Research Fund.
The research activities described in this paper have been co-funded by the European Union - NextGenerationEU within PRIN 2022 project n.20229YBSAN - 
Globular clusters in cosmological simulations and in lensed fields: from their birth to the present epoch.
\end{acknowledgements}

%%%%%%%%%%%%%%%%%%%%%%%%%%%%%%%%%%%%%%%%%%%%%%%%%%



%\bibliographystyle{mnras}

\bibliographystyle{aa}

%%%%%%%%%%%%%%%%%%%%%%%%%%%%%%%%%%%%%%%%%%%%%%%%%%

%%%%%%%%%%%%%%%%% APPENDICES %%%%%%%%%%%%%%%%%%%%%

\begin{appendix}
\section{Additional table}
\label{Appendix A}

%%%Best-fit values table

%\begin{sidewaystable*}
%\footnotesize
\begin{table}[!h]
%\centering
\caption{Best-fit kinematic parameters for the MPs in the target GCs.}
\label{tab:mp_kin} 
\scriptsize
\begin{tabular}{lcccccccccccccc}
\hline \hline 
Cluster & $\sigma_{0}^{LOS}$ & $\sigma_{0}^{RAD}$ & $\sigma_{0}^{TAN}$ &  $A_{rot}^{LOS}$ & $R_{peak}^{LOS}$ & $PA_{0}$      & $A_{rot}^{TAN}$ & $R_{peak}^{TAN}$ & $A_{rot}^{3D}$ & $i$           &    $\theta_0$     &     $\beta_{\infty}$ & $\epsilon$ & POP \\
        &   [km/s]           &     [km/s]         &     [km/s]         &       [km/s]     &    [$\arcsec$]   &  [$^{\circ}$] &     [km/s]      &    [$\arcsec$]   &    [km/s]      & [$^{\circ}$]  &    [$^{\circ}$]   &                      &      &     \\
\hline
NGC 104        &  $11.4^{+1.7}_{-1.2}$ & $10.5^{+0.9}_{-0.7}$ & $13.7^{+1.8}_{-1.2}$  &  $-2.6^{+0.6}_{-0.7}$ & $195.9^{+70.6}_{-120.5}$& $53^{+7}_{-5}$    & $4.7^{+1.6}_{-1.5}$  & $248.2^{+25.9}_{-28.7}$ & $-6.5^{0.33}_{0.31}$  &    $30^{+7}_{-6}$   & $69^{+11}_{-12} $ &$-2.3^{+1.8}_{-1.8}$ & $0.15^{+0.01}_{-0.01}$ &  FP   \\
               &  $9.9^{+1.3}_{-0.7}$ & $11.6^{+0.8}_{-0.6}$ & $10.1^{+0.7}_{-0.6}$  &  $-3.0^{+0.3}_{-0.3}$ & $283.0^{+58.1}_{-67.1}$  &$30^{+8}_{-6}$    & $6.3^{+0.8}_{-0.8}$  & $220.0^{+22.1}_{-22.8}$ &$-8.0^{+0.37}_{0.36}$    &  $32^{+7}_{-6}$   & $57^{+9}_{-9}   $ &$ 0.3^{+0.1}_{-0.1}$ & $0.26^{+0.01}_{-0.01}$  &  SP	\\							 
NGC 288        &  $2.5^{+0.3}_{-0.2}$ & $3.5^{+0.5}_{-0.3}$ & $2.9^{+0.2}_{-0.2}$  & $0.4^{+0.3}_{-0.3}$ & $161.4^{+27.1}_{-32.8}$ &     $-2^{+44}_{-24}$  & $-1.8^{+1.2}_{-1.1}$ & $637.4^{+227.1}_{-219.8}$ & $0.5^{+0.4}_{-0.2}$  & $41^{+23}_{-32}$ & $332^{+18}_{-18}$ &$-0.8^{+2.8}_{-2.8}$  & $0.19^{+0.08}_{-0.04}$ &  FP	\\
               &  $3.3^{+0.7}_{-0.5}$ & $4.8^{+2.4}_{-1.2}$ & $5.6^{+2.4}_{-1.2}$  & $0.5^{+0.3}_{-0.3}$ & $151.6^{+32.4}_{-29.2}$ &     $1^{+28}_{-36}$   & $-1.6^{+1.5}_{-1.4}$ & $470.6^{+242.4}_{-182.2}$ & $1.5^{+0.3}_{-0.3}$  & $18^{+26}_{-38}$ & $35^{+24}_{-25} $ &$1.4^{+2.1}_{-2.8}$   & $0.07^{+0.03}_{-0.03}$ &  SP	\\
NGC 1261       &  $4.0^{+1.2}_{-0.6}$ & $4.8^{+5.8}_{-1.8}$ & $4.1^{+1.5}_{-0.8}$  & $-1.8^{+1.0}_{-0.9}$ & $46.4^{+97.3}_{-35.1}$ &     $-31^{+20}_{-20}$ &$0.4^{+1.4}_{-1.1}$ & $195.4^{+103.8}_{-90.7}$ & $3.0^{+1.4}_{-1.4}$      &$78^{+12}_{-12}$ & $43^{+27}_{-28} $ &$-0.7^{+3.4}_{-2.8}$  & $0.24^{+0.13}_{-0.05}$ &  FP   \\
               &  $3.3^{+0.6}_{-0.4}$ & $5.8^{+5.4}_{-1.8}$ & $3.3^{+1.6}_{-0.9}$  & $-1.7^{+0.6}_{-0.7}$ & $50.8^{+24.4}_{-19.7}$  &    $-31^{+19}_{-17}$ & $0.8^{+0.9}_{-0.7}$ & $110.0^{+138.2}_{-101.6}$  & $2.5^{+1.3}_{-1.3}$   &$63^{+7}_{-6}$   & $47^{+23}_{-23} $ &$0.2^{+2.3}_{-3.2}$  & $0.27^{+0.03}_{-0.02}$ &  SP   \\	
NGC 1904       &  $3.8^{+0.7}_{-0.4}$ & $4.2^{+1.1}_{-0.5}$ & $5.5^{+1.8}_{-1.0}$  & $1.6^{+0.7}_{-0.8}$ & $31.9^{+39.3}_{-15.2}$  &     $-8^{+15}_{-4}$   & $-1.1^{+1.6}_{-1.8}$ & $101.4^{+71.2}_{-49.2}$ & $1.5^{+1.2}_{-1.2}$ &    $51^{+18}_{-19}$ & $29^{+31}_{-29} $ &$0.3^{+2.7}_{-3.0}$  & $0.06^{+0.05}_{-0.04}$ &  FP   \\
               &  $4.4^{+1.0}_{-0.6}$ & $5.5^{+2.4}_{-1.0}$ & $5.8^{+1.6}_{-1.2}$  & $0.9^{+0.6}_{-0.5}$ & $58.8^{+50.7}_{-38.6}$  &     $24^{+20}_{-30}$  &$-2.1^{+2.1}_{-1.9}$ & $259.5^{+183.6}_{-113.6}$ & $2.5^{+0.7}_{-0.7}$ &   $23^{+26}_{-23}$ & $36^{+26}_{-26} $ &$0.2^{+2.6}_{-2.6}$   & $0.06^{+0.03}_{-0.03}$ &  SP \\     
NGC 3201       &  $4.2^{+0.4}_{-0.3}$ & $3.7^{+0.3}_{-0.2}$ & $3.6^{+0.3}_{-0.2}$  & $0.3^{+0.8}_{-1.0}$ & $315.3^{+126.1}_{-159.6}$ &   $20^{+52}_{-91}$  &  $-0.6^{+1.0}_{-0.9}$ & $345.3^{+78.1}_{-72.6}$ & $1.2^{+0.5}_{-0.5}$ &    $44^{+22}_{-21}$ & $331^{+36}_{-37}$ &$0.3^{+2.2}_{-2.5}$  & $0.30^{+0.02}_{-0.04}$ &  FP   \\
               &  $3.6^{+0.2}_{-0.2}$ & $3.6^{+0.3}_{-0.2}$ & $3.5^{+0.4}_{-0.3}$  & $0.1^{+0.9}_{-0.6}$ & $262.2^{+164.9}_{-168.2}$ &   $-13^{+73}_{-50}$ &  $-0.5^{+0.7}_{-0.8}$ & $440.1^{+151.9}_{-92.2}$ & $1.0^{+0.5}_{-0.5}$ &   $41^{+28}_{-21}$ & $345^{+32}_{-37}$ &$0.7^{+0.3}_{-0.2}$ & $0.10^{+0.02}_{-0.02}$  &  SP   \\			   
NGC 5272       &  $7.5^{+1.4}_{-0.9}$ & $8.4^{+1.2}_{-0.8}$ & $6.9^{+0.8}_{-0.6}$  & $0.6^{+0.5}_{-0.5}$ & $120.4^{+100.0}_{-100.2}$ &   $46^{+24}_{-62}$  &  $2.3^{+1.6}_{-1.8}$ & $225.9^{+88.2}_{-78.8}$ & $1.0^{+0.5}_{-0.5}$ &     $33^{+27}_{-27}$ & $78^{+18}_{-18}$  &$1.7^{+2.1}_{-1.7}$   & $0.26^{+0.08}_{-0.08}$  &  FP   \\
               &  $6.8^{+3.8}_{-1.2}$ & $7.3^{+0.9}_{-0.7}$ & $7.6^{+1.7}_{-0.9}$  & $0.8^{+0.4}_{-0.4}$ & $450.0^{+93.2}_{-238.3}$  &   $34^{+18}_{-80}$  &  $3.1^{+1.6}_{-1.2}$ & $89^{+54.9}_{-57.4}$ &  $2.0^{+0.4}_{-0.4}$&        $56^{+21}_{-22}$ & $61^{+16}_{-18}$  &$1.0^{+1.9}_{-0.9}$ & $0.22^{+0.04}_{-0.07}$ &  SP   \\     
NGC 5904       &  $7.8^{+2.2}_{-1.2}$ & $7.4^{+1.3}_{-0.9}$ & $7.0^{+0.9}_{-0.7}$  & $2.6^{+0.4}_{-0.4}$ & $234.7^{+110.3}_{-68.7}$ &    $-27^{+9}_{-10}$  &  $-3.7^{+1.9}_{-1.6}$ & $148.0^{+42.3}_{-33.7}$ & $4.1^{+0.4}_{-0.5}$ &    $49^{+6}_{-6}$   & $27^{+16}_{-18}$  &$-2.3^{+1.5}_{-1.7}$ & $0.36^{+0.02}_{-0.03}$ &  FP   \\
               &  $7.2^{+1.5}_{-0.9}$ & $6.7^{+0.9}_{-0.6}$ & $7.5^{+0.9}_{-0.7}$  & $2.1^{+0.3}_{-0.3}$ & $237.6^{+92.5}_{-62.5}$ &     $-58^{+6}_{-7}$   &   $-3.2^{+1.6}_{-1.4}$ & $240.3^{+68.5}_{-77.5}$ & $5.5^{+0.5}_{-0.5}$ &   $40^{+6}_{-6}$   & $25^{+18}_{-19}$  &$0.5^{+2.0}_{-1.9}$    & $0.36^{+0.02}_{-0.02}$ &  SP   \\     
NGC 5927       &  $5.3^{+1.1}_{-0.6}$ & $6.8^{+0.3}_{-0.3}$ & $6.7^{+0.4}_{-0.4}$  & $-0.8^{+0.8}_{-0.5}$ & $305.3^{+387.3}_{-167.8}$ &  $-32^{+50}_{-30}$ &  $0.5^{+1.6}_{-1.6}$ & $56.6^{+125.3}_{-35.6}$ & $1.0^{+0.8}_{-0.1}$  &    $37^{+35}_{-29}$ & $305^{+27}_{-27}$ &$-9.8^{+3.3}_{-3.2}$ & $0.01^{+0.01}_{-0.01}$  &  FP   \\
               &  $5.6^{+0.7}_{-0.5}$ & $7.0^{+0.3}_{-0.3}$ & $6.0^{+1.4}_{-0.8}$  & $-0.2^{+1.8}_{-1.8}$ & $320.4^{+396.1}_{-180.0}$ &  $-6^{+80}_{-68}$  &  $0.6^{+1.8}_{-1.8}$ & $451.1^{+244.2}_{-214.6}$   & $1.5^{+0.8}_{-0.8}$&  $58^{+33}_{-29}$ & $315^{+36}_{-38}$ &$0.5^{+0.1}_{-0.2}$ & $0.06^{+0.03}_{-0.03}$ &  SP   \\     
NGC 5986       &  $8.8^{+9.9}_{-2.9}$ & $8.3^{+3.9}_{-2.6}$ & $8.5^{+2.2}_{-2.2}$  & $1.1^{+1.1}_{-1.0}$ & $288.7^{+134.7}_{-188.1}$ &   $14^{+40}_{-55}$  & $-1.9^{+2.8}_{-1.8}$ & $210.7^{+120.1}_{-110.7}$ & $1.9^{+0.5}_{-0.5}$ &  $80^{+35}_{-32}$ & $312^{+19}_{-19}$  &$3.0^{+2.8}_{-1.9}$  & $0.35^{+0.06}_{-0.06}$ &  FP   \\
               &  $8.3^{+3.5}_{-1.7}$ & $9.6^{+1.5}_{-1.0}$ & $8.5^{+0.8}_{-0.6}$  & $1.6^{+1.5}_{-2.5}$ & $309.4^{+120.3}_{-154.6}$ &   $-40^{+88}_{-29}$ &  $-3.0^{+2.2}_{-2.5}$ & $40.5^{+54.7}_{-38.4}$ & $1.2^{+0.5}_{-0.4}$    & $80^{+35}_{-33}$ & $308^{+19}_{-19}$  &$-4.2^{+4.4}_{-6.1}$  & $0.35^{+0.06}_{-0.06}$   &  SP   \\   
NGC 6093       & $11.7^{+0.6}_{-0.5}$ & $11.5^{+0.6}_{-0.5}$ & $11.7^{+0.7}_{-0.6}$ & $2.5^{+1.5}_{-1.4}$& $70.8^{+16.8}_{-13.8}$&         $76^{+30}_{-39}$  &  $2.5^{+1.5}_{-1.4}$ & $70.8^{+16.8}_{-13.8}$ & $2.1^{+0.5}_{-0.5}$        & $117^{+34}_{-34}$& $122^{+9}_{-9}$   &$-0.88^{+0.58}_{-0.93}$ & $0.11^{+0.04}_{-0.04}$ & FP \\	  
               & $10.8^{+0.7}_{-0.6}$ & $11.0^{+0.4}_{-0.5}$  & $10.8^{+0.7}_{-0.6}$ & $2.1^{+1.2}_{-1.2}$ & $80.8^{+12.5}_{-11.9}$ &        $59^{+39}_{-37}$  &  $2.1^{+1.2}_{-1.2}$ & $70.8^{+12.5}_{-11.9}$ & $1.6^{+0.5}_{-0.5}$      &   $122^{+28}_{-29}$& $125^{+13}_{-7}$  &$-0.72^{+0.87}_{-1.02}$ & $0.11^{+0.04}_{-0.04}$ &  SP \\   
NGC 6171       &  $3.4^{+0.6}_{-0.4}$ & $3.5^{+0.3}_{-0.3}$ & $3.9^{+0.4}_{-0.3}$  & $0.4^{+0.4}_{-0.3}$ & $332.0^{+436.2}_{-205.2}$ &   $4^{+60}_{-67}$   &  $-0.7^{+1.4}_{-2.0}$ & $91.7^{+136.3}_{-89.8}$ & $-1.0^{+0.5}_{-0.5}$ &  $30^{+55}_{-58}$ & $23^{+41}_{-43}$  &$-2.0^{+3.3}_{-6.6}$ & $0.11^{+0.02}_{-0.03}$ &  FP   \\
               &  $3.5^{+0.8}_{-0.4}$ & $4.7^{+1.2}_{-0.7}$ & $4.2^{+0.8}_{-0.4}$  & $0.4^{+0.3}_{-0.2}$ & $171.5^{+204.9}_{-81.6}$ &    $16^{+50}_{-73}$  &  $-0.8^{+1.2}_{-1.3}$ & $233.6^{+158.3}_{-136.8}$ & $-0.5^{+0.6}_{-0.6}$ &$40^{+54}_{-55}$&  $34^{+41}_{-43}$  &$-4.2^{+4.0}_{-6.4}$ & $0.05^{+0.03}_{-0.03}$   &  SP   \\	      
NGC 6205       &  $8.8^{+2.5}_{-1.9}$ & $8.7^{+1.9}_{-1.6}$ & $8.8^{+1.7}_{-1.4}$  & $2.4^{+2.7}_{-1.7}$ & $174.0^{+132.1}_{-102.2}$ &   $53^{+48}_{-85}$  &  $2.4^{+1.9}_{-2.2}$ & $174^{+111}_{-89}$ & $-1.6^{+0.7}_{-0.7}$    &     $157^{+16}_{-19}$& $85^{+10}_{-12}$  &$0.02^{+0.98}_{-1.32}$ & $0.34^{+0.04}_{-0.05}$ &   FP \\	   
               &  $6.1^{+0.9}_{-0.6}$  & $6.3^{+1.2}_{-1.2}$ & $6.1^{+0.9}_{-0.6}$ & $2.9^{+0.9}_{-0.9}$ & $113.6^{+90.6}_{-46.7}$   &   $22^{+37}_{-45}$  & $2.9^{+0.9}_{-0.9}$  & $113.6^{+90.6}_{-46.7}$ & $-2.5^{+0.7}_{-0.7}$   &  $173^{+16}_{-19}$& $78^{+10}_{-12}$  &$2.86^{+0.57}_{-0.76}$ & $0.36^{+0.06}_{-0.05}$  &  SP  \\	 
NGC 6362       &  $3.5^{+0.4}_{-0.3}$  & $3.4^{+0.5}_{-0.5}$ & $3.5^{+0.3}_{-0.4}$ & $0.6^{+0.1}_{-0.1}$ & $37.3^{+12.7}_{-22.8}$ &        $102^{+40}_{-40}$ & $0.7^{+0.1}_{-0.1}$ & $17.3^{+2.7}_{-2.8}$ & $0.4^{+0.3}_{-0.3}$ &        $62^{+34}_{-33}$ & $140^{+38}_{-36}$ &$-0.11^{+1.58}_{-1.39}$ & $0.02^{+0.02}_{-0.02}$ & FP \\
               &  $3.6^{+0.4}_{-0.3}$  & $3.7^{+0.3}_{-0.3}$ & $3.6^{+0.3}_{-0.3}$ & $1.2^{+0.2}_{-0.1}$ & $32.3^{+22.5}_{-24.8}$ &        $120^{+36}_{-36}$ & $1.2^{+0.2}_{-0.1}$ & $17.3^{+2.7}_{-2.8}$ & $0.7^{+0.1}_{-0.2}$ &        $49^{+21}_{-24}$ & $114^{+27}_{-28}$ &$0.53^{+1.25}_{-1.22}$  & $0.11^{+0.02}_{-0.02}$ & SP  \\     
NGC 6254       &  $5.4^{+0.6}_{-0.4}$ & $4.8^{+0.6}_{-0.4}$ & $5.3^{+0.6}_{-0.4}$  & $0.7^{+0.5}_{-0.4}$ & $649.7^{+292.2}_{-440.5}$ &   $-1^{+64}_{-58}$  & $-0.3^{+1.2}_{-1.1}$ & $465.7^{+397.2}_{-326.5}$ & $1.5^{+0.5}_{-0.5}$ &  $81^{+26}_{-27}$ & $317^{+33}_{-31}$ &$-2.9^{+2.4}_{-5.8}$  & $0.11^{+0.05}_{-0.03}$ &  FP \\
               &  $4.8^{+0.5}_{-0.3}$ & $5.2^{+0.6}_{-0.4}$ & $5.1^{+0.8}_{-0.5}$  & $0.5^{+0.6}_{-0.4}$ & $531.6^{+379.9}_{-391.3}$ &   $3^{+53}_{-57}$   & $-0.1^{+1.0}_{-1.0}$ & $418.6^{+428.9}_{-315.3}$ & $1.0^{+0.5}_{-0.5}$ &  $73^{+23}_{-27} $& $25^{+24}_{-21}$  &$3.1^{+3.1}_{-2.1}$   & $0.09^{+0.03}_{-0.03}$ &  SP \\			 
NGC 6496       &  $2.9^{+0.6}_{-0.4}$ & $4.2^{+0.6}_{-0.4}$ & $4.3^{+1.6}_{-1.1}$  & $0.9^{+1.0}_{-0.6}$ & $307.7^{+167.1}_{-258.7}$ &   $18^{+44}_{-75}$  &  $-0.4^{+1.2}_{-0.4}$ & $150.7^{+120.1}_{-110.7}$ & $0.5^{+0.6}_{-0.6}$ & $124^{+48}_{-55}$& $28^{+43}_{-41}$  &$5.6^{+5.9}_{-4.6}$ & $0.34^{+0.06}_{-0.06}$  &  FP   \\
               &  $2.4^{+0.6}_{-0.4}$ & $3.9^{+0.9}_{-0.6}$ & $3.4^{+1.0}_{-0.7}$  & $1.0^{+1.1}_{-0.7}$ & $315^{+160.7}_{-237.4}$ &     $9^{+61}_{-74}$   & $0.80^{+0.8}_{-0.7}$ & $65.5^{+54.7}_{-38.4}$ & $1.0^{+0.7}_{-0.7}$  &    $129^{+51}_{-52}$& $23^{+44}_{-21}$  &$0.4^{+5.3}_{-7.1}$ & $0.34^{+0.06}_{-0.05}$  &  SP   \\			    
NGC 6723       &  $4.6^{+0.8}_{-0.5}$ & $6.2^{+1.5}_{-0.8}$ & $6.8^{+1.9}_{-1.4}$  & $0.7^{+0.5}_{-0.4}$ & $161.0^{+269.1}_{-76.3}$ &    $-28^{+53}_{-25}$ &  $-0.1^{+1.1}_{-1.9}$ & $277.2^{+251.5}_{-182.5}$ & $2.4^{+0.4}_{-0.4}$  &$1^{+37}_{-35}$  & $340^{+23}_{-22}$ & $0.5^{+2.6}_{-2.3}$  & $0.26^{+0.04}_{-0.05}$ &  FP   \\
               &  $4.6^{+0.9}_{-0.5}$ & $5.3^{+0.3}_{-0.3}$ & $5.9^{+0.3}_{-0.3}$  & $0.5^{+0.5}_{-0.3}$ & $155.1^{+275.6}_{-75.4}$ &    $-19^{+72}_{-39}$ &  $3.3^{+2}_{-2}$ & $72.9^{+27.5}_{-27.5}$ & $1.5^{+0.4}_{-0.5}$ &         $80^{+27}_{-27}$ & $350^{+21}_{-21}$ &$-1.8^{+1.9}_{-2.0}$   & $0.05^{+0.03}_{-0.03}$&  SP   \\		
\hline                 
\end{tabular}
\end{table}
%\end{sidewaystable}

\clearpage
\section{Incompleteness effects}
\label{Appendix B}
We constrained the possible impact of the kinematic samples' size and of their (radially dependent) incompleteness (mainly caused by the intrinsically limited allocation efficiency of multi-object spectrographs) on the results obtained in this work by using the dynamical simulations described in Sections 5 and 6. 

In detail, we estimated the completeness ($C$) of the observed kinematic samples and its radial variation as the ratio between the number of RGB stars detected in the photometric catalogs and those with \texttt{LOS} RVs and/or PMs measures within concentric radial annuli at different cluster-centric distances. While we acknowledge that these estimates represent a lower-limit to the real incompleteness, as photometric catalogs are not fully complete, we note that our targets are RGB stars, which are among the brightest stars in GC CMDs and therefore they are only moderately affected by incompleteness even when ground-based catalogs are adopted.

We then extracted randomly from the simulations sub-samples of stars with similar sizes as the observed ones for a large number of times. We applied the derived completeness curves to these sub-samples to make them as similar as possible to the observed catalogs. Finally we run the same kinematic analysis described in Section~3.
We find that, while the limited sample sizes and incompleteness have an impact on the overall noise of the observed kinematic profiles and as a consequence on the uncertainties associated to the derived parameters, they do not have a significant impact on the final results.  

\section{Global kinematics and comparison with the literature}
\label{Appendix C}
\begin{figure}[ht!]
\centering
\includegraphics[width=\columnwidth]{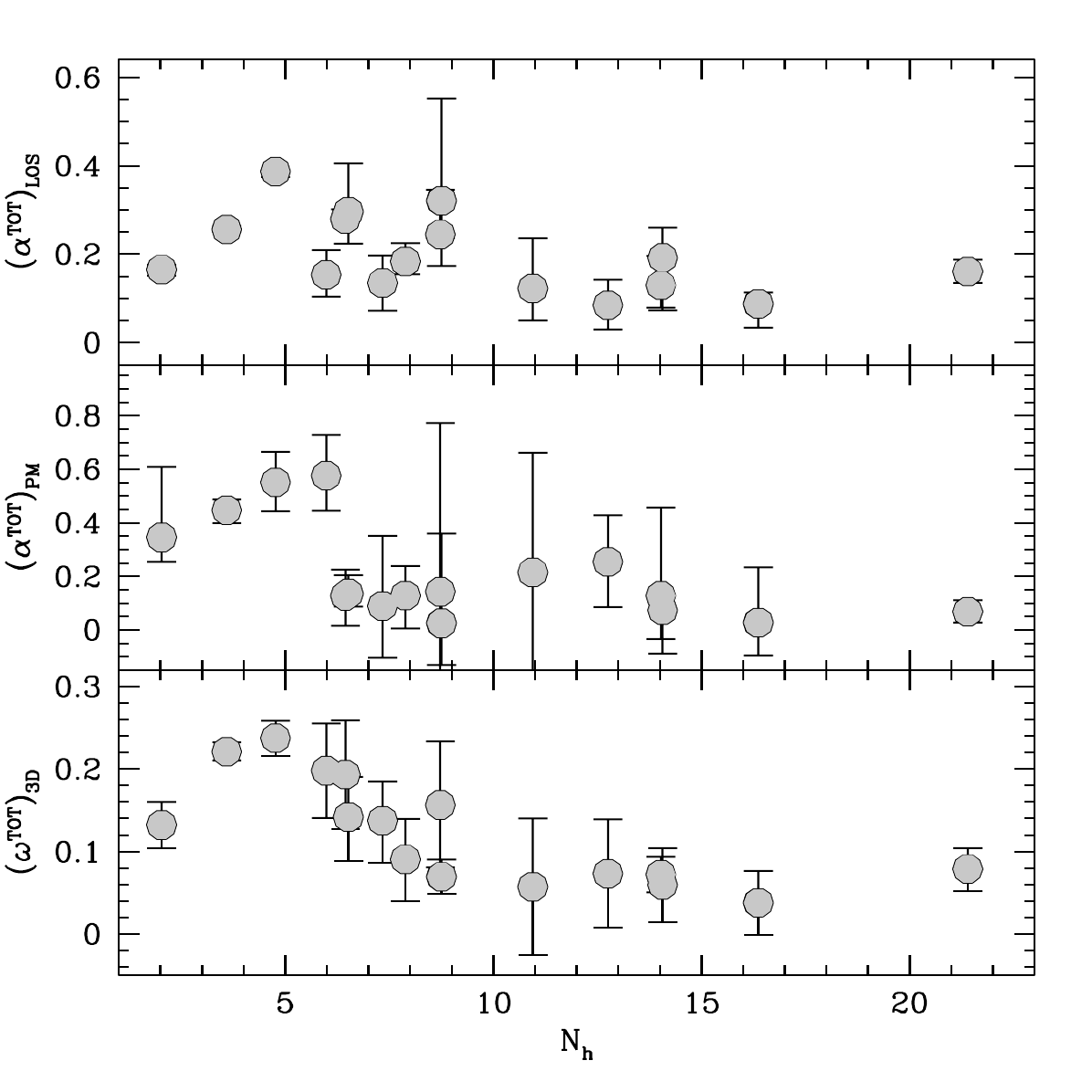}
\caption{Distribution of the three rotation parameters defined in this work for the \texttt{LOS}, \texttt{PM} and 3D velocity components, as a function of the dynamical age ($N_{\rm h}$) for the total population (TOT) of GCs in our sample.}
\label{fig:rot_tot}
\end{figure}

Figure~\ref{fig:rot_tot} shows the distribution of the $\alpha$ values derived
for the entire population along both the \texttt{LOS} and \texttt{TAN} velocity components, and of the $\omega_{3D}^{\rm TOT}$, as a function of $N_{\rm h}$. 
As expected, both $\alpha^{\rm TOT}$ (for both \texttt{LOS} and \texttt{TAN}) and the $\omega_{3D}^{\rm TOT}$ distributions show pretty clear anti-correlations with $N_{\rm h}$. In fact, while dynamically young GCs attain larger values of rotation parameters, with the only significant 
exception being M~3, the rotation strength progressively decreases as dynamical evolution proceeds. By performing a Spearman rank correlation test we find that such anti-correlations have a significance of $\sim98\%$ for the \texttt{LOS} and $>99.9\%$ for both the \texttt{TAN} and the 3D analysis, with the 3D case being the most significant. 
In general, these results are in very good agreement with previous analysis (e.g., \citealt{kamann18,sollima19}) and they further strengthen the conclusion that 
the present-day cluster rotation is the relic of that imprinted at the epoch of cluster formation, which has been then progressively dissipated via two-body relaxation.

\begin{figure}[ht!]
\centering
\includegraphics[width=\columnwidth]{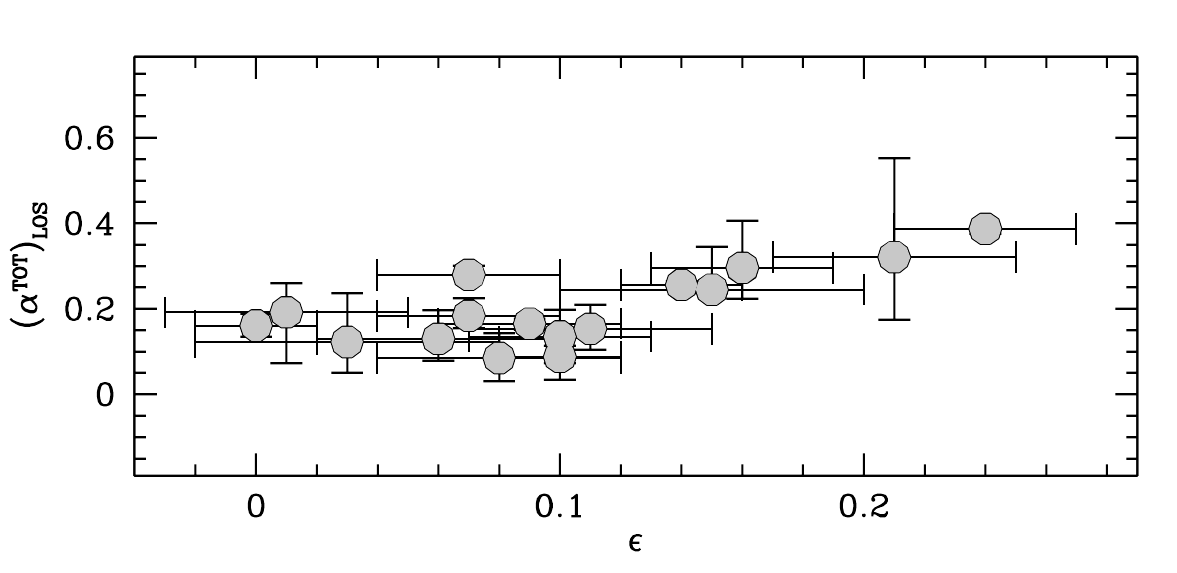}
\caption{Distribution of the ($\alpha^{\rm TOT})_{\rm LOS}$ parameter for the total population as a function of the best-fit ellipticity values derived as described in Section~4.}
\label{fig:ell_tot}
\end{figure}

Figure~\ref{fig:ell_tot} shows the distribution of $\alpha_{\rm LOS}^{\rm TOT}$ with the cluster ellipticity obtained as described in Section~3.
As expected (see discussion and references in Section~5.2), a nice correlation between rotation and ellipticity is observed also for the total population in very good agreement with previous findings by \citet{fabricius14,kamann18}.

\begin{figure*}
\includegraphics[width=0.95\textwidth]{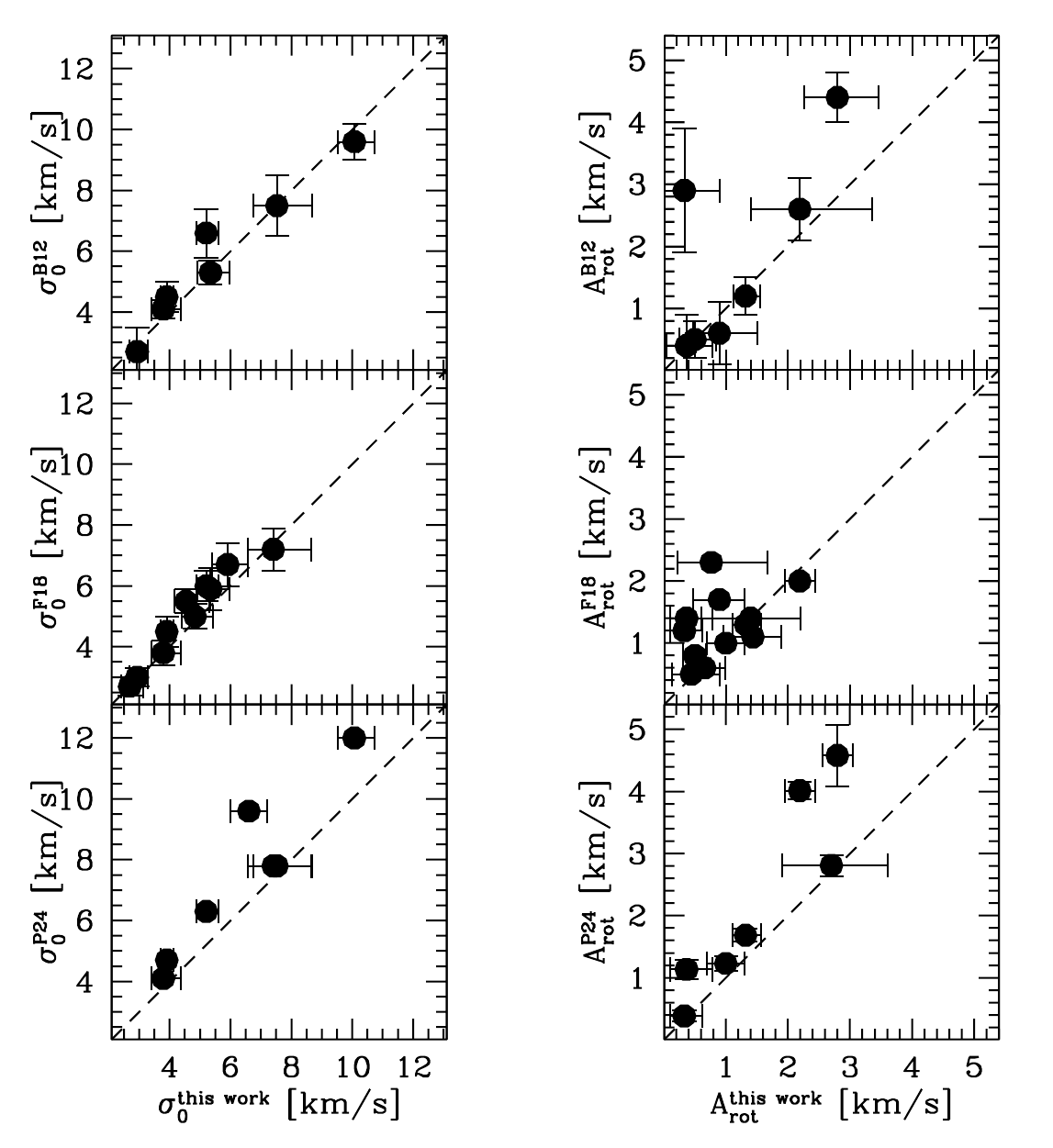}
\caption{Comparison between the best-fit $\sigma_0$ and $A_{\rm rot}$ values obtained for the TOT sample for the clusters in common between the present work and \citet{bellazzini12} -- B12, \citet{ferraro18} -- F18 and \citet{petralia24} -- P24.}
\label{fig:comparison}
\end{figure*}

While the focus of this work is on the MP kinematics, nevertheless it is useful to compare the results obtained for the TOT population with those largely available in the literature to have an indication about the general performance of the adopted approach and data-sets. 

\begin{figure}[!ht]
\centering
\includegraphics[scale=0.3]{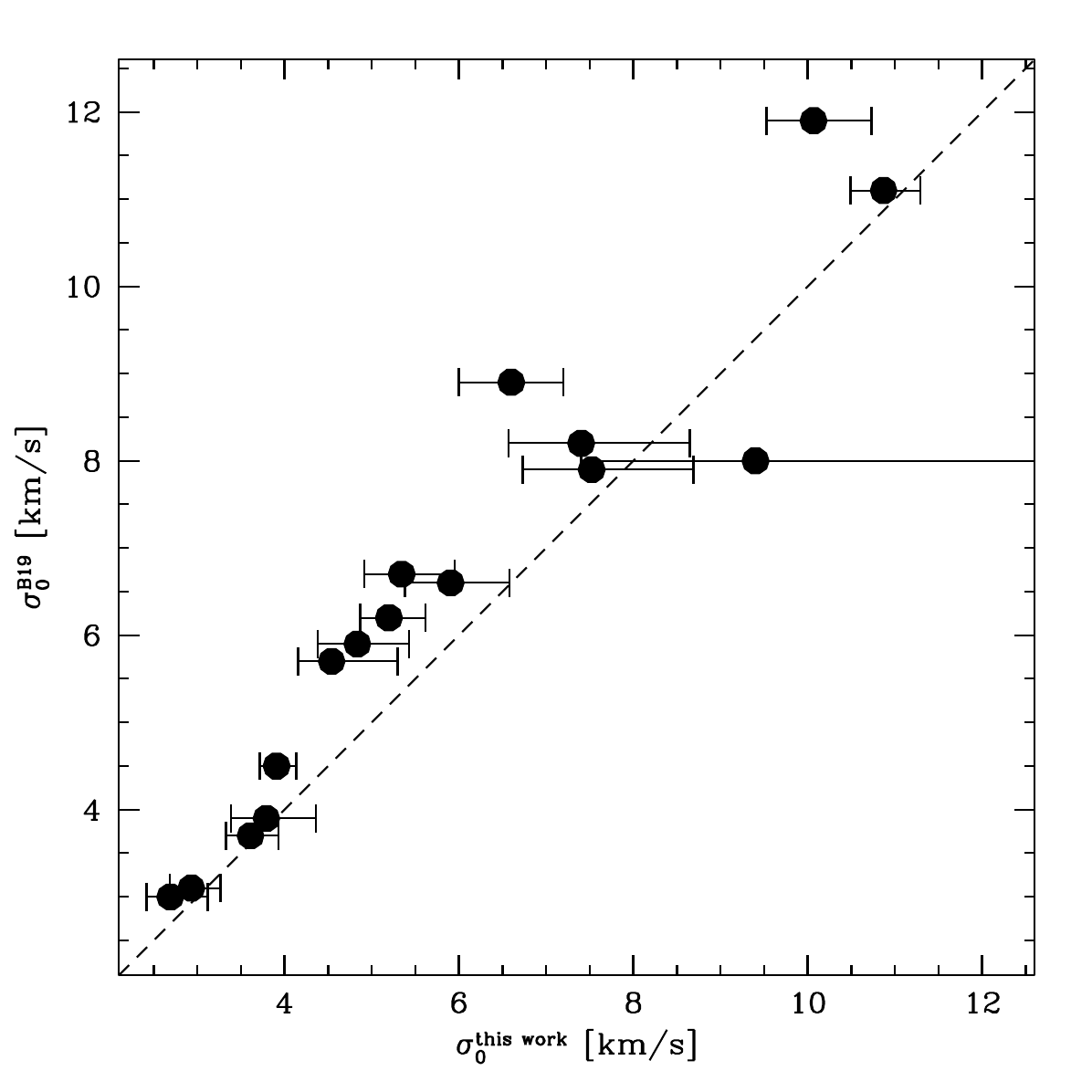}
\includegraphics[scale=0.3]{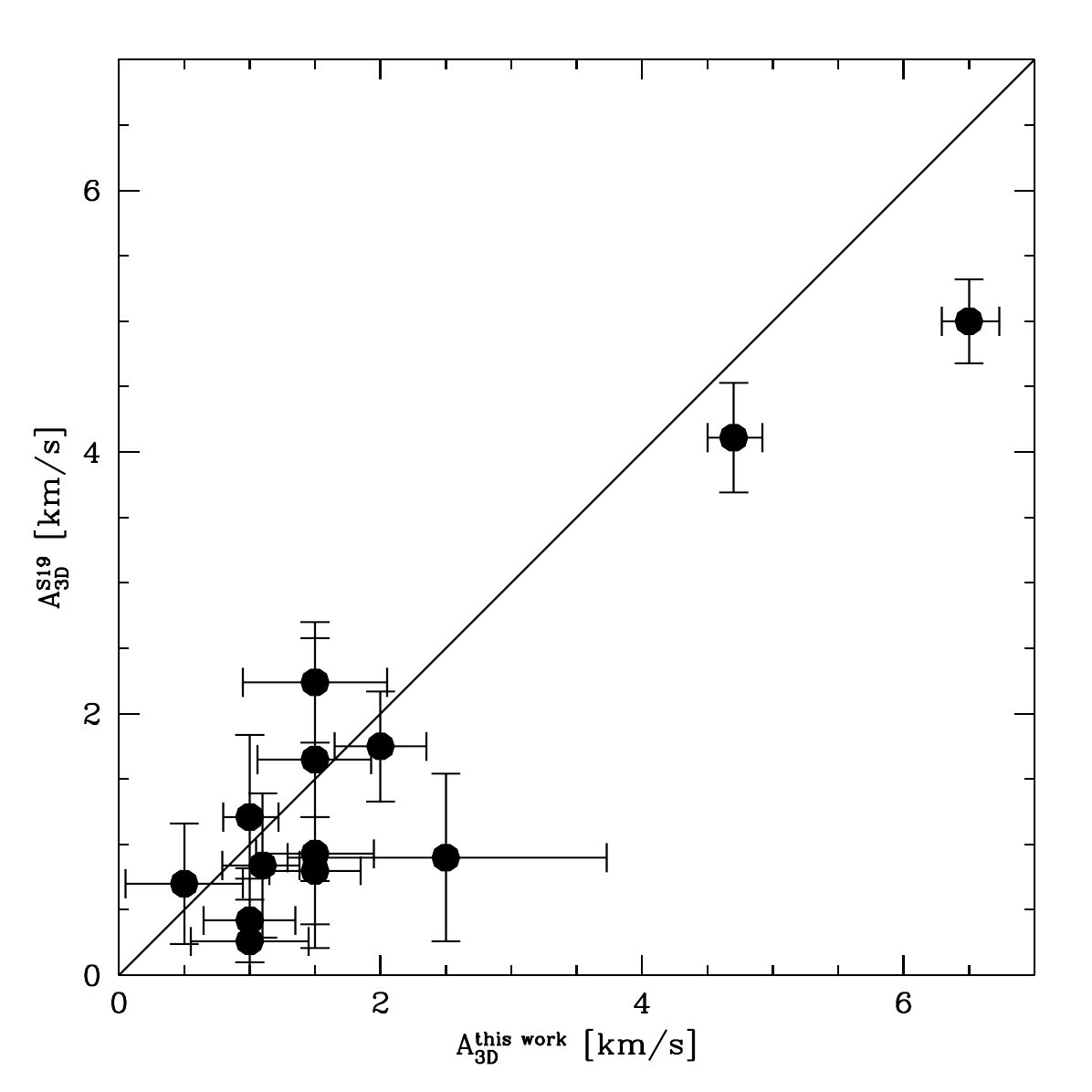}
\caption{Left panel: comparison with the best-fit $\sigma_0$ values obtained by \citet{baumgardt19} -- B19 -- for the total population of GCs in common with the present work. Right-panel: one-to-one comparison of the 3D rotation amplitude values $A_{\rm 3D}$ found by \citet{sollima19} -- S19 -- and the present analysis.}
\label{fig:comparisonB19_S19}
\end{figure}

Detailed one-to-one comparisons with recent results obtained in the literature \citep{bellazzini12,ferraro18,lanzoni18a,lanzoni18b,baumgardt19,sollima19} for the TOT population are shown in Figures~\ref{fig:comparison} and ~\ref{fig:comparisonB19_S19}. 
In general, a quite good agreement is observed with all the compilations considered here.

Our sample has 6 GCs in common with \citet{bellazzini12}. A nice match is observed both in terms of $\sigma_0$ and $A_{\rm rot}$ (top row of Figure~\ref{fig:comparison}) with the only exception being NGC~6171 for which \citet{bellazzini12} finds a rotation amplitude 4-5 times larger than the one obtained in this work. Given the estimate by \citet{bellazzini12}, NGC~6171 would be a very fast rotator, with $A_{rot}/\sigma_0\sim0.7$. However, it is important to note, that the sample of \texttt{LOS} RVs used by \citet{bellazzini12} for this cluster includes only 
31 stars in total, resulting the smallest sample of \texttt{LOS} RVs  
in their analysis. Here we sample the kinematic profile of NGC~6171 with 184 \texttt{LOS} RVs (see Table~\ref{tab:gc}).
We note also that NGC~6171 results to have a significantly smaller rotation amplitude ($1.2$ km/s) than what found by \citet{bellazzini12} in the analysis by \citet{ferraro18} and it is classified as non rotator by \citet{sollima19}

As for the comparison with results by \citet{ferraro18}, we stress that while the spectroscopic sample is largely similar, the adopted kinematic analysis (both the discrete and continuous ones) is significantly different for the rotation study in particular (see \citealt{ferraro18} for details). Hence, it is not surprising that while the derived central velocity dispersion values are in excellent agreement for the entire sample (middle row of Figure~\ref{fig:comparison}), the distribution of differences for $A_{\rm rot}$ is more scattered, while still showing a reasonable match within the errors. In this case, the most significant discrepancy is observed for NGC~5927, for which \citet{ferraro18} derived $A_{\rm rot}=2.3$ km/s, while we find 
$A_{rot}=0.76^{+0.90}_{-0.54}$ km/s. For this cluster also \citet{sollima19} derived a low probability of rotation.

In the bottom row of Figure~\ref{fig:comparison} we compare the results of this work with those recently obtained by \citet{petralia24} by using APOGEE spectra for a sample of Galactic GCs. For $A_{\rm rot}$ we use the semi-amplitude of the $A_{\rm fit}$ values reported in their work. A reasonable overall agreement is observed also in this case for the clusters in common, however 47~Tuc and NGC~5904 result to have larger central velocity dispersion values and peak of rotation than in our work.

Finally, as shown in Figure~\ref{fig:comparisonB19_S19} (left panel) a reasonably good match is also found with the $\sigma_0$ estimates by \citet{baumgardt19}.
Among the clusters in common with \citet{sollima19}, the only significantly discrepant result is that of 47~Tuc, which results to have a $\sim25\%$ larger rotation in this work. However, we note that in this case, as for the entire sample, both the $i$ and $\theta_0$ values are in very good agreement. In this respect, it is also interesting to highlight the nice match in terms of both the observed rotation amplitude and angles of the 3D rotation of 47~Tuc obtained in this work and those inferred by means of a detailed comparison between HST PMs and theoretical models of rotating clusters by \citet{bellini17}.

\end{appendix}


%%%%%%%%%%%%%%%%%%%%%%%%%%%%%%%%%%%%%%%%%%%%%%%%%%

% Don't change these lines

\end{document}